\documentclass[journal]{IEEEtran}

\ifCLASSINFOpdf
  \usepackage[pdftex]{graphicx}
  \graphicspath{{./Figures/}{./photos/}}
  \DeclareGraphicsExtensions{.pdf,.png,.jpg}
\else
  \usepackage[dvips]{graphicx}
  \graphicspath{{./Figures/}{./photos/}}
  \DeclareGraphicsExtensions{.pdf,.png}
\fi

\usepackage{amsmath}
\ifCLASSOPTIONcompsoc
 \usepackage[caption=false,font=normalsize,labelfont=sf,textfont=sf]{subfig}
\else
 \usepackage[caption=false,font=footnotesize]{subfig}
\fi
\usepackage{dblfloatfix}
\usepackage{url}

\usepackage{amsfonts}

\usepackage{flushend}

\usepackage{enumitem}

\usepackage{multirow}

\usepackage{cases}

\newcommand\given[1][]{\:#1\vert\:}

\newcommand{\epc}{Euler--Poincar\'e}

\usepackage{algorithm}
\usepackage{algorithmic}
\usepackage{siunitx}

\DeclareSIUnit\year{yr}

\makeatletter
\newcommand\fs@norules{\def\@fs@cfont{\bfseries}\let\@fs@capt\floatc@ruled
  \def\@fs@pre{}%
  \def\@fs@post{}%
  \def\@fs@mid{\kern3pt}%
  \let\@fs@iftopcapt\iftrue}
\makeatother
\floatstyle{norules}
\restylefloat{algorithm}

\begin{document}

%
\title{Bone Adaptation as a Geometric Flow}
%
%
%


\author{
  Bryce~A.~Besler, 
  Tannis~D.~Kemp,
  Nils~D.~Forkert, 
  Steven~K.~Boyd

\thanks{This work was supported by the Natural Sciences and Engineering Research Council (NSERC) of Canada, grant RGPIN-2019-4135.}
\thanks{B.A. Besler and T. D. Kemp are in the McCaig Institute for Bone and Joint Health, University of Calgary Canada. N.D. Forkert is with the Department of Radiology and the Hotchkiss Brain Institute, University of Calgary, Canada. S.K. Boyd is with the Department of Radiology and McCaig Institute for Bone and Joint Health, University of Calgary Canada e-mail: skboyd@ucalgary.ca}

\thanks{Manuscript received November 18, 2020; revised TODO}}

%
%

\markboth{}%
{Besler \MakeLowercase{\textit{et al.}}: Constructing High-Order Signed Distance Maps from Computed Tomography}
%



\maketitle

\begin{abstract}
  This paper presents bone adaptation as a geometric flow.
  The proposed method is based on two assumptions: first, that the bone surface is smooth (not fractal) permitting the definition of a tangent plane and, second, that the interface between marrow and bone tissue phases is orientable.
  This permits the analysis of bone adaptation using the well-developed mathematics of geometric flows and the numerical techniques of the level set method.
  Most importantly, topological changes such as holes forming in plates and rods disconnecting can be treated formally and simulated naturally.
  First, the relationship between biological theories of bone adaptation and the mathematical object describing geometric flow is described.
  This is termed the adaptation function, $F$, and is the multi-scale link described by Frost's Utah paradigm between cellular dynamics and bone structure.
  Second, a model of age-related bone loss termed curvature-based bone adaptation is presented.
  Using previous literature, it is shown that curvature-based bone adaptation is the limiting continuous equation of simulated bone atrophy, a discrete model of bone aging.
  Interestingly, the parameters of the model can be defined in such a way that the flow is volume-preserving.
  This implies that bone health can in principle change in ways that fundamentally cannot be measured by areal or volumetric bone mineral density, requiring structure-level imaging.
  Third, a numerical method is described and two \textit{in silico} experiments are performed demonstrating the non-volume-preserving and volume-preserving cases.
  Taken together, recognition of bone adaptation as a geometric flow permits the recruitment of mathematical and numerical developments over the last 50 years to understanding and describing the complex surface of bone.
\end{abstract}


%
\IEEEpeerreviewmaketitle

\section{Introduction}
\label{sec:introduction}
The concept of bone adaptation as a geometric flow is presented.
The work presented here is a significant extension of a conference proceeding~\cite{besler2018bone}.
Particularly, an artifact of signed distance transforms of sampled signals has been identified~\cite{besler2020artifacts} and solved in the case of computed tomography images of biphasic materials~\cite{besler2021constructing}.
The mathematics have been expanded significantly to tightly link the model to the theory of geometric flows.

\section{Adaptation as a Geometric Flow}
It is assumed that bone changes occur at the interface of marrow and bone tissue.
As a consequence of this claim, with an assumption of smoothness, many statements can be made about the underlying dynamics.
Specifically, they can be modeled as a geometric flow where the flow rate has a historic and important meaning in the theory of bone adaptation.

\subsection{Biology of Bone Adaptation}
Bone adaptation occurs fundamentally at the surface~\cite{frost1969tetracycline,frost1987bone,frost2000utah}.
This is in opposition to ontogenesis~\cite{berendsen2015bone} and indirect fracture healing~\cite{marsell2011biology} where endochondral ossification is replacing cartilage or intramembranous ossification is occurring directly from sheets of mesenchymal connective tissue.
Functional adaptation refers to adaptation controlled principally through mechanics, typically coordinated by the osteocyte summarizable by a biological set point theory termed the mechanostat~\cite{frost1987bone}.
Adaptation is separated into modeling (motion of the periosteal and endosteal surfaces through surface drifts) and remodeling (changes in cortical and trabecular bone through coordinated cell action)~\cite{frost1987bone}.
The unit of remodeling is the basic multicellular unit (BMU) consisting of osteoblasts, osteoclasts, osteocytes, and other cells coordinated through cellular dynamics.
Remodeling is coordinated differently in the lacunae of trabecular bone~\cite{raggatt2010cellular} and the osteon's of cortical bone~\cite{eriksen2010cellular}.
Furthermore, a distinction is made between changes in shape (external remodeling) and changes in the material properties (internal remodeling)~\cite{beaupre1990approachtheory}.
The principle concept is that adaptation occurs on the surface, which presupposed the existence of a surface as opposed to a density field and that adaptation can be modeled as a change in this surface over time.

\subsection{The Bone Surface}
\label{subsec:the_bone_surface}
Let bone be described by a density field $\rho : \Omega \rightarrow \mathbb{R}^+$ defined on a domain $\Omega \subset \mathbb{R}^3$.
It is assumed that bone is a biphasic material consisting of the marrow phase and the bone tissue phase, the domain is a union of the two phases $\Omega = \Omega_\text{Marrow} \cup \Omega_\text{Tissue}$ where $\Omega_\text{Marrow}$ and $\Omega_\text{Tissue}$ are the marrow and bone tissue components in the field, respectively.
Importantly, since the bone is a biphasic material, its interface can be described as an orientable surface:
\begin{equation}
  \mathcal{C} : \mathbb{R}^2 \rightarrow \mathbb{R}^3
\end{equation}
where $\mathcal{C}$ is the surface.
The consequence of having an orientable surface is that there is a defined inside and outside, so that a volume can be defined and area elements oriented.

One additional claim is made that the surface is locally smooth, permitting differentiation.
Since the surface is differentiable, the area is finite, a tangent plane can be defined at each point on the surface, and principle, mean, and Gaussian curvature defined.
This constraint will be relaxed in Section~\ref{subsec:adaptation_as_a_geometric_flow} to permit topological changes during adaptation.

\begin{figure*}[t]
  \centering
  \begin{tabular}{ccc}
    \subfloat[A Rod Resorbing]{
      \includegraphics[width=0.3\linewidth]{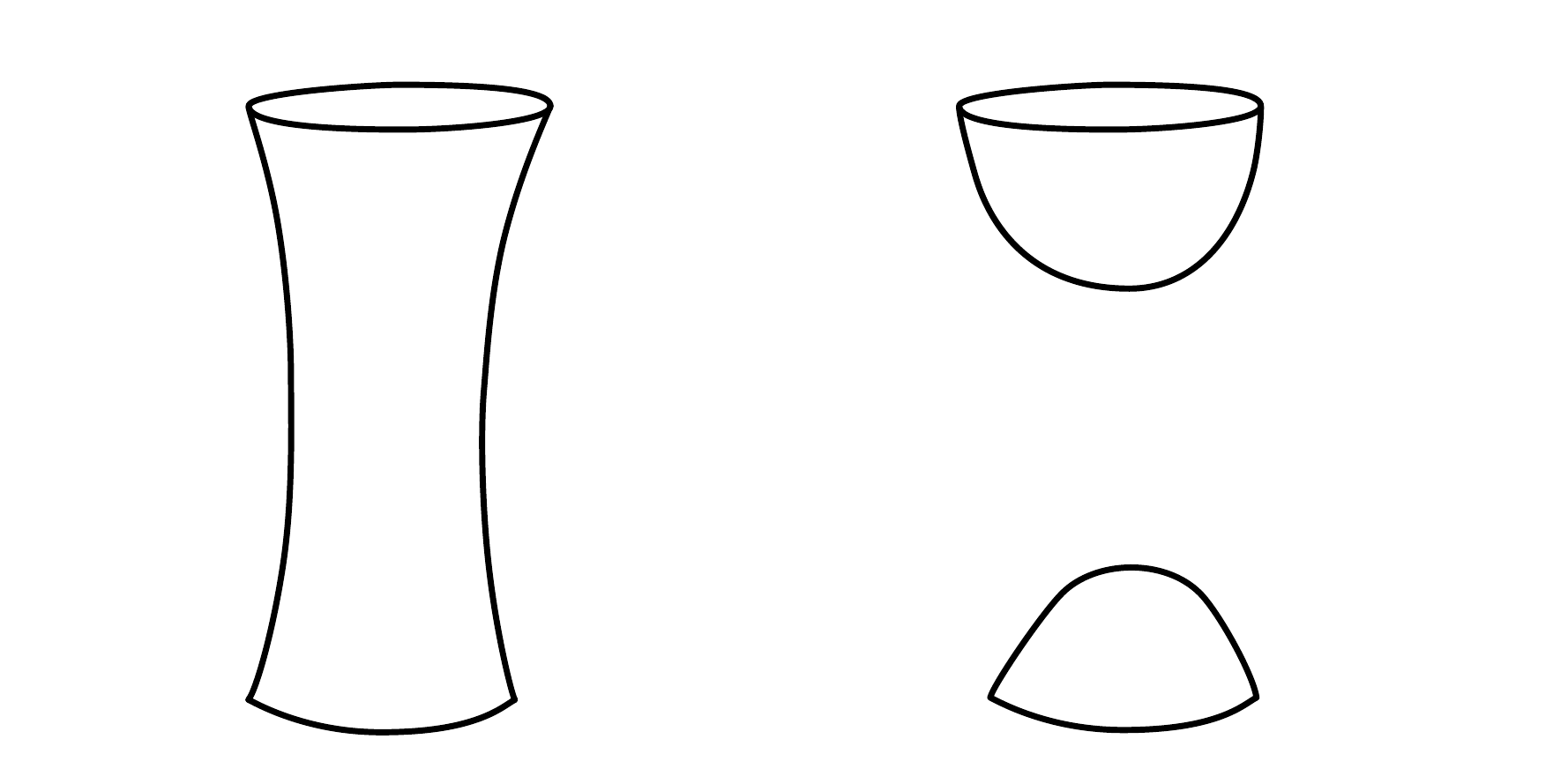}%
      \label{fig:idealized:rod}
    } &
    \subfloat[A Plate Forming a Hole]{
      \includegraphics[width=0.3\linewidth]{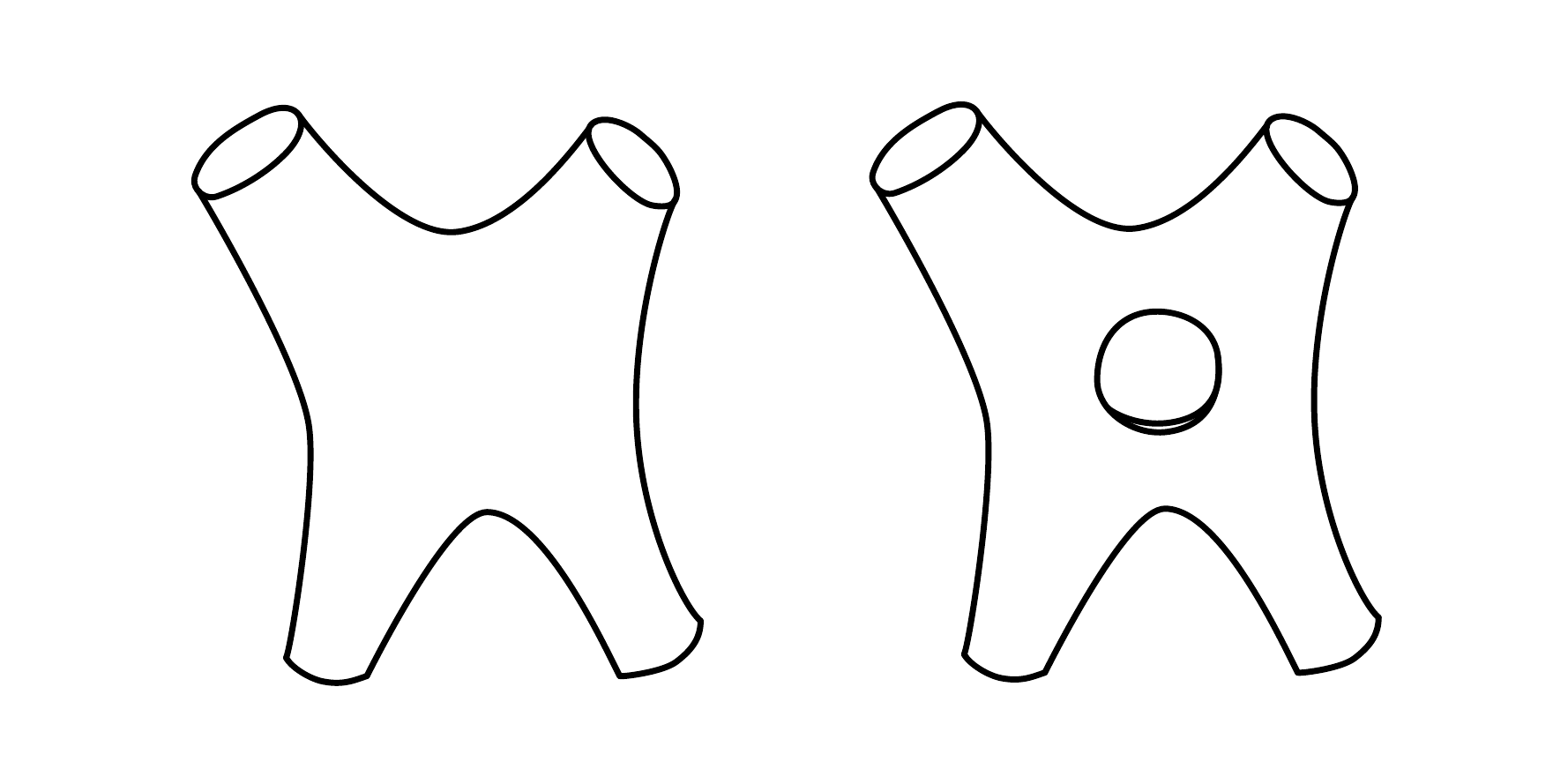}%
      \label{fig:idealized:plate}
    }&
    \subfloat[Periosteal Drift]{
      \includegraphics[width=0.3\linewidth]{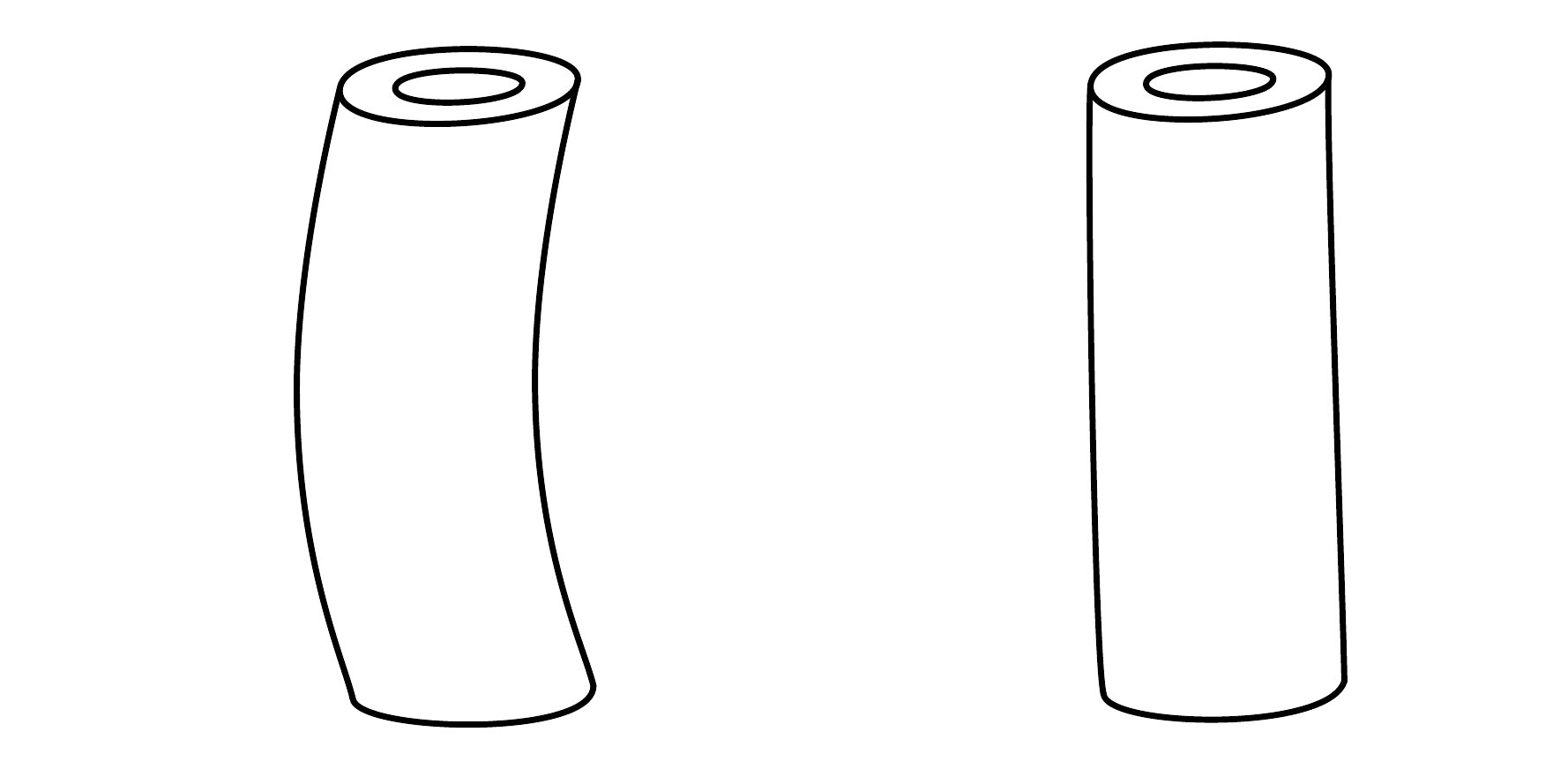}%
      \label{fig:idealized:drift}
    }
  \end{tabular}
  \caption{Idealized surface changes in bone structure. (\ref{fig:idealized:rod}) Rods can resorb, changing topology; (\ref{fig:idealized:plate}) plates can form holes, causing rod-to-plate transition; and (\ref{fig:idealized:drift}) periosteal drift can change the gross morphometry.}
  \label{fig:idealized}
\end{figure*}

In contrast to differential geometry~\cite{kreyszig2019differential}, one could define bone using the theory of fractal geometry~\cite{mandelbrot1982fractal}.
Fractal dimension is a well established morphometric parameter of trabecular bone~\cite{majumdar1993application,fazzalari1996fractal} defined from fractal geometry and there are strong arguments that bone has fractal properties~\cite{geraets2000fractal}.
The coastline paradox is the quintessential natural experiment about the origin of fractals, where the measured length of a coastline depends on the size of the ruler you measure it with.
The paradox is that the area continues to increase as the ruler decreases, all the way down to the scale of an atom, leading to one having a curve with finite volume but infinite area.
In bone, the perimeter is replaced with surface area and the ruler with the resolution of the imager.
While this behavior is confirmed at \textit{in vivo} resolutions~\cite{fazzalari1996fractal}, it is not clear if this trend would continue \textit{ad infinitum}.
At some resolution --- say, the $\SI{100}{\nano\meter}$ scale --- the measured area is assumed to stabilize.
Scanning electron microscopy confirms relatively smooth surfaces, albeit some surface roughness~\cite{boyde1986scanning}.
While this theory holds that the surface is smooth, no claims are made on properties of the surface not exhibiting fractal-like behavior in the form of power laws, such as the distribution of pore size~\cite{jorgenson2015comparison}.

\subsection{Adaptation as a Geometric Flow}
\label{subsec:adaptation_as_a_geometric_flow}
Having established the bone surface as an orientable, smooth surface, attention is placed on how to move the surface in time.
This leads directly to an extrinsic geometric flow:
\begin{equation}
  \label{eqn:curve_normal}
  \mathcal{C}_t = FN
\end{equation}
where $\cdot_t$ is a partial derivative in time, $F$ is a rate of surface growth that varies spatially across the surface, and $N$ is the normal of the surface.
This equation captures motion only in the normal direction along the surface since tangential motion leads to a reparameterization of the surface and not a change in its geometry.
With an initial surface, this presents bone adaptation as an initial value (Cauchy) problem:
\begin{subequations}
  \begin{numcases}{}\label{eqn:ivp}
    \mathcal{C}_t = FN & \label{eqn:ivp:pde}\\
    \mathcal{C}(x,0) = \mathcal{C}_0 & \label{eqn:ivp:surface}
  \end{numcases}
\end{subequations}
where $C_0$ is the initial bone surface.
Such a formulation has been used extensively in computational physics~\cite{sussman1994level} and active contours~\cite{kass1988snakes}.

Classic results of extrinsic geometric flows follow naturally from presentation of the Cauchy problem~\cite{huisken1984flow,gage1986heat,sethian1999level}.
Changes in topology can occur as rods disconnect or holes form in plates.
In finite time, the curve can develop sharp corners, which are continuous but not smooth, requiring special treatment through the theory of viscous solutions~\cite{crandall1983viscosity}.
A complete contrast and comparison of geometric flows and their relation to bone adaptation is beyond the scope of this work and is an area of future interest.

\subsection{The Adaptation Function, $F$}
The principle consequence of considering bone as a geometric flow is that the quantity $F$, combined with the initial surface, completely describes the adaptation of bone.
Due to its importance, we term the quantity $F$ the \textit{adaptation function}.

\subsubsection{$F$ in Frost's mechanostat}
$F$ is exactly the mechanostat graph of Carter~\cite{carter1984mechanical} and Frost~\cite{frost2001wolff}.
Remodeling, lamellar bone drifts, and woven bone drifts are summarized in a single graph where changes in bone density (or surface) are a function of local mechanical strain. 
It is a summary of the BMU and changes spatially across the surface of bone.
This paradigm has been used extensively to develop computational models of bone adaptation~\cite{adachi1997simulation,adachi2001trabecular,huiskes2000effects,schulte2011invivo,kameo2014modeling}.
Attempts have been made to measure $F$ experimentally and establish the presence of lazy zones~\cite{christen2014bone}.


\subsubsection{$F$ in Dynamical Systems}
One paradigm for understanding skeletal health is to treat the basic multicellular unit (BMU) as a dynamical system.
Hormones circulating in the blood stream (PTH, calcitonin, calcitriol, estrogens, etc.) and cytokines expressed locally (OPG, RANK, Wnts, TGF-$\beta$, etc.) control the rate of bone formation and resorption.
Importantly, these substrates change the temporal dynamics of other substances, say through the down regulation of PTH as calcium leaves the bone and enters the blood stream or through the modified expression of RANK-L from osteoblasts, which leads to nonlinear effects in neighboring cells.
Dynamical systems based on nonlinear partial differential equations have been used extensively to model coupled system.
At the cellular level, dynamical models have explained paradoxes in experimental research~\cite{komarova2003mathematical} as well as predicting the response of bone to different cytokines~\cite{lemaire2004modeling,pivonka2010mathematical}.
The adaptation function, $F$, is a continuous summary of the activity of discrete osteoblasts and osteoclasts, similar in concept to diffusion being a continuous summary of quantized particles moving from Brownian motion.
In this paradigm, the adaptation function is the link between cellular- and tissue-scale dynamics~\cite{gerhard2009insilico,webster2001insilico}.

\subsubsection{$F$ in Cellular Automata}
Dynamical systems lead immediately to the paradigm of cellular automata (although not equivalent, taken here in spirit with complex adaptive systems and agent based modeling)~\cite{neumann1951general,langton1986studying,wolfram2002new}.
The central concept of cellular automata is that complex behaviors can emerge from iterating simple rules, emerging in a way non-obvious by studying the rules in isolation.
Such models have been used for predicting cortical remodeling~\cite{buenzli2012investigation}, posing bone adaptation as a topological optimization problem~\cite{tovar2005bone}, and simulating fracture healing~\cite{tourolle2019micro}.
Viewed as agents, osteoblast and osteoclast cells can be thought to interact through simple rules, being defined by cytokines, local loading, and genetics.
The system adapts through time, causing the spatial pattern of bone to emerge.
In contrast to dynamical systems, this is a computational paradigm of bone adaptation.
As in dynamical systems, the adaptation function $F$ is a summary of these local agents.

\subsection{Remarks on Internal and External Models}
There is a rich history in the development of models of bone adaptation.
We differentiate those models that make the two-phase assumption from those that do not with Carter's terminology of internal remodeling~\cite{beaupre1990approach,huiskes2000effects} and external remodeling~\cite{cowin1985functional,adachi1997simulation,schulte2013strain}, respectively.
There are models that make use of both~\cite{hart1984mathematical}.
Equation~\ref{eqn:curve_normal} is the central equation for external remodeling while the central equation for internal remodeling is:
\begin{equation}
  \rho_t = F
\end{equation}
where $\rho$ is the density field.

In both cases, the adaptation function $F$ remains the central quantity of investigation.
In internal remodeling methods, $F$ has dimensions of density per unit time ([Mass][Length]\textsuperscript{-3}[Time]\textsuperscript{-1}) while in external remodeling methods, $F$ has dimensions of length per unit time ([Length][Time]\textsuperscript{-1}).
While these models are tightly coupled through $F$, their assumption on the density field have different consequences.
For instance, internal remodeling can have a hole form in the center of a trabecula, not connected to the marrow space.
Unless one relaxes the reinitialization condition (Section~\ref{subsubsec:reinitialization}), this cannot occur in external remodeling.
A consequence of this is that the analytic models of Weinans~\cite{weinans1992behavior} and Cowin~\cite{cowin1981bone} are not equivalent.

\section{Curvature-Based Bone Adaptation}
We now present a specific model of bone adaptation for age-related bone loss.
The model is based heavily on a prior model of age-related bone loss called simulated bone atrophy (SIBA)~\cite{muller1996analysis,mueller1997biomechanical,pistoia2003mechanical,muller2005long}.
Using prior literature, it will be demonstrated that SIBA simulates mean curvature flow, providing a strong link to this geometric model.

\subsection{The Model}
Curvature-based bone adaptation models age-related changes in bone loss as a summation of advection of the surface and mean curvature flow.
This gives the adaptation function:
\begin{equation}
  \label{eqn:cbba}
  F = a - b \kappa
\end{equation}
where $a$ is the advection constant with dimensions [Length][Time]\textsuperscript{-1}, $b$  is the curvature constant with dimensions [Length]\textsuperscript{2}[Time]\textsuperscript{-1}, and $\kappa$ is the mean curvature with dimensions [Length]\textsuperscript{-1}.
While $a$ can take on any value, $b$ can only take on positive values.
Negative values of $b$ are consistent with inverse mean curvature flow, which is not defined for flat surfaces and becomes numerically unstable for non-flat surfaces.
This is a two-parameter model defining a spatially varying adaptation function that depends only on the local geometry.

The intuition behind Equation~\ref{eqn:cbba} is the same as in SIBA.
Thin connections in the bone will resorb first not only because they are smaller, but because they have much higher curvature.
Holes can form in plates causing plate-to-rod transition.
Generally, changes occur on the surface, the trabecular bone surface erodes, and the rate at which bone changes varies across the surface.

An important definition is when $F=0$ across the surface, as the bone will stop adapting.
Rearranging Equation~\ref{eqn:cbba}, the stopping condition can be found:
\begin{equation}
  F = b(\langle \kappa \rangle - \kappa)
\end{equation}
where $\langle \kappa \rangle = a/b$ is the average mean curvature across the surface.
The bone will stop adapting when it is a surface of constant mean curvature, $\kappa = \langle \kappa \rangle$ everywhere.
Such an equation is seen in the Young-Laplace equation~\cite{laplace1805traite,young1805essay} describing soap films, surface tension, and capillary rise.

A closed form solution for curvature-based bone adaptation is difficult even for simple analytic surfaces.
A solution is given for the sphere~\cite{besler2020sphere} and has been long known for the cases when $a=0$ or $b=0$.
However, simple analytic surfaces are in someway unfaithful for developing intuitions on the equation.

\subsection{Relation to Minimal Surfaces}
A minimal surface is a surface of smallest area given a constraint.
Equivalently, as mean curvature is the first variation of area, a minimal surface will have a mean curvature of zero everywhere.
The first of such surfaces were Euler's catenoid and helicod.
Later, Schwarz and Neovius described periodic minimal surfaces that extended infinitely.
Schoen later classified these surfaces and discovered the gyroid~\cite{schoen1970infinite}
Finally, surfaces of non-zero and spatially varying constant mean curvature were explored by Chopp and Sethian~\cite{chopp1991computing,chopp1993flow} using level set methods.
Such surfaces have been used for designing scaffolds for tissue engineering~\cite{kapfer2011minimal} and lightweight but strong structures~\cite{zhang2018energy}.

\subsection{Relation to SIBA}
The relationship between curvature-based bone adaptation and simulated bone atrophy (SIBA) is now described.
SIBA works on binary images of bone where the bone tissue phase is the foreground object and the marrow phase is the background object.
A finite support Gaussian filtration is used to blur the object followed by a threshold to rebinarize the image.
Physical interpretations were given to the variance of the Gaussian blur and threshold value based on osteoblast efficiency and resorption depth. 
The key methodological novelty of SIBA was that it naturally handled bone changing topology, where rods could resorb and plats could form holes.

The intuitive similarities between SIBA and curvature-based bone adaptation is that Gaussian blurring moves the surface in a way that resembles mean curvature flow.
Local changes on the bone surface are dependent on the magnitude and sign of the local mean curvature.
As such, an advection term is used to model thresholding and a mean curvature term is used to model the Gaussian blurring.
However, this link can be made concrete, and is done so now.
The main result is that if a threshold of $0.5$ is used in SIBA, the method is equivalent to mean curvature flow in the limit as the product of kernel size and epoch time goes to zero.

\subsubsection{Advection --- Threshold Link}
\label{subsubsec:siba:advection}
The link between advection and thresholding is derived following a previous derivation for Gaussian smoothed surfaces (Appendix of ~\cite{besler2021morph}).
Consider a binary image, $I$, blurred with some Gaussian filter, $G_\sigma$:
\begin{equation}
  \label{eqn:siba_blur}
  J = I * G_\sigma
\end{equation}
where $*$ is convolution and $J$ is the resulting grayscale image.
We seek to understand how far the surface moves given a threshold $T$ of $J$.
The intensity is normalized by forcing the binary image $I$ to take on values of $0$ or $1$.

The curvature-dependence of the problem is removed by considering a blurring much smaller than the local mean curvature ($\sigma \ll |H|$) such that the surface is near flat (relative to the smoothing).
Then, we can reduce the three-dimensional problem to a one-dimensional problem by considering the line along the surface normal, since blurring does not change the surface in the tangent plane.
We define $I$ to be a unit step (Heaviside) function $\theta(x)$ with zero crossing at $x = 0$.
The unit step is substituted into Equation~\ref{eqn:siba_blur} and the location of the $T$ level set is computed.
\begin{eqnarray}
  J(x) &=& \theta(x) * G_\sigma \\
  T &=& \frac{1}{2}\left(1 + \text{erf}\left(\frac{x}{\sqrt{2}\sigma}\right)\right) \\
  \label{eqn:siba:advection}
  x &=& \sqrt{2}\sigma \text{erf}^{-1}\left(2T-1\right)
\end{eqnarray}
where $\text{erf}(\cdot)$ is the error function:
\begin{equation}
  \text{erf}(z) = \frac{2}{\sqrt{\pi}} \int_0^z e^{-t^2}dt
\end{equation}
and $\text{erf}^{-1}$ the inverse error function.
Note that the inverse error function is unique if $|z| < 1$, which corresponds to thresholds inside the dynamic range $[0, 1]$ specified for the binary image.
Furthermore, $\text{erf}^{-1}(0) = 0$, corresponding to no change in the surface if $T=0.5$.

Equation~\ref{eqn:siba:advection} gives an estimation of the distance the surface travels in flat areas for a given threshold $T$.
Dividing this distance by the sample time, which is the inverse of activation frequency (AF) in SIBA, gives an estimate of the corresponding advection constant:
\begin{equation}
  \label{eqn:siba:a}
  a = \sqrt{2}\sigma \text{erf}^{-1}\left(2T-1\right) \text{AF}
\end{equation}

\subsubsection{Mean Curvature --- Gaussian Filtration Link}
\label{subsubsec:siba:mean}
The relationship follows in two steps.
First, consider the heat flow of the image:
\begin{subequations}
  \begin{numcases}{}\label{eqn:heat}
    I_\tau = \alpha \Delta I & $\text{on }\Omega \times (0, \infty)$ \label{eqn:heat:pde}\\
    I(x,0) = I_0 & $\text{on }\Omega \times 0$ \label{eqn:heat:iv}
  \end{numcases}
\end{subequations}
where $I_0$ is the original binary image, $\tau \in [0, \infty)$ is a time-like parameter, $\alpha$ is thermal diffusivity typically set to unity, and $\Delta = \nabla^2$ is the Laplacian operator.
This defines a one-parameter family of images where the time-like parameter captures the scale of objects in the image~\cite{witkin1984scale,koenderink1984structure}.
It is well known that the solution of the heat equation is Gaussian convolution:
\begin{equation}
  I(x,\tau) = I_0(x) * G_\tau(x)
\end{equation}
where $2\alpha\tau = \sigma^2$ in Equation~\ref{eqn:siba_blur}.
As such, one can work with $I(x,\tau)$ equivalently to the Gaussian blurring.
The problem now reduces to comparing heat flow of the image to mean curvature flow of the surface.

Realizing the link between Gaussian convolution and heat flow, SIBA is exactly the same as the BMO (Bence-Merriman-Osher) algorithm in computational physics for simulating mean curvature flow, except a threshold different from $0.5$ is selected~\cite{merriman1992diffusion}.
BMO simulates mean curvature flow by blurring a binary image using the heat equation and rebinarizing the field with a threshold at $0.5$.
Evans (Theorem 5.1, \cite{evans1993convergence}) proved that if $u$ is the viscous solution from mean curvature flow and $I(x,\tau)$ the solution from the diffusion equation, the two methods are equivalent in the limit of small $\tau$.
The consequence is that for small values of $\sigma/\text{AF}$ and with $T=0.5$, SIBA simulates mean curvature flow.

Following the methods of the BMO algorithm~\cite{merriman1992diffusion} in spherical coordinates, the mean curvature constant can be estimated as twice the thermal diffusivity constant:
\begin{equation}
  b = 2 \alpha
\end{equation}
Note that $b$ and $\alpha$ have the same dimensions, [Length]\textsuperscript{2}[Time]\textsuperscript{-1}.
Substituting into the relationship between $\alpha$ and $\sigma$:
\begin{equation}
  \label{eqn:siba:b}
  b = \frac{\sigma^2}{\tau}
\end{equation}
where $\tau$ can be estimated as the inverse of activation frequency, as in the case of advection.
The factors-of-two cancel because the surface is embedded in three dimensions.
Finally, one can estimate the target mean curvature in SIBA by dividing Equation~\ref{eqn:siba:a} by Equation~\ref{eqn:siba:b}:
\begin{equation}
  \langle \kappa \rangle = \frac{\sqrt{2} \text{erf}^{-1}\left(2T-1\right)}{\sigma}
\end{equation}
Noting the restriction in Section~\ref{subsubsec:adv_dis}, the algorithm will stop before this condition is met.

\subsubsection{Limitations on these Similarities}
It should be noted that the derivations in Section~\ref{subsubsec:siba:advection} and \ref{subsubsec:siba:mean} are approximate and not exact.
Additional analysis is needed to derive bounds and convergence orders.
Differences will arise because SIBA is compositional (Gaussin blur then thresholding) while the proposed model is additive (summing advection and mean curvature flow).
Finally, SIBA was designed to be implemented in discrete space with a finite support Gaussian filter, causing differences to this continuous model.

\subsubsection{Advantages and Disadvantages}
\label{subsubsec:adv_dis}
There are three primary advantages to curvature-based bone adaptation over simulated bone atrophy.
The first advantage is a high-order representation of the bone surface.
Since the surface is represented as a signed distance transform, gradients are available, and sub-voxel shifts can be tracked over time.
In SIBA, the bone is represented by a binary image that is continually blurred and rebinarized.
This inherently limits the representation to first order accurate, $\mathcal{O}(h)$, since information on the derivatives is lost by binarization.
Furthermore, the bone surface can never traverse through voxel edges in very flat surfaces before being rebinarized, causing the surface to stop advecting artificially early (see Section 4 of~\cite{merriman1992diffusion}).
This problem has an intricate link to anti-aliasing filters in computer graphics~\cite{blinn1989jim,besler2020artifacts}.
The second advantage is well-defined mathematics for geometric flows.
This gives us principled methods of understanding when the flow stops, if the flow minimizes an energy functional, and how the area and volume change with the flow.
Further refinement of the link between dynamic histomorphometry~\cite{frost1969tetracycline,parfitt1987bone,schulte2011vivo} and geometric flows~\cite{sapiro2006geometric} as well as formulating energy functionals for bone adaptation~\cite{huiskes2000if} is an exciting future direction.
The final advantage is that building load driven adaptation models based on the binary representation has the distinct disadvantage of requiring blurring to establish normal vectors~\cite{schulte2013strain}.
The Gaussian filtration has an inherent trade-off where large blurs are needed to prevent quantization of the normal vector while small blurs are preferred to limit structural changes.
Additionally, this causes an implicit mean curvature flow on top of the expected load driven adaptation, presenting difficulties in model validation.

\begin{figure*}
  \centering
  \begin{tabular}{cccc}
    \subfloat[parameterization]{
      \includegraphics[width=0.21\linewidth]{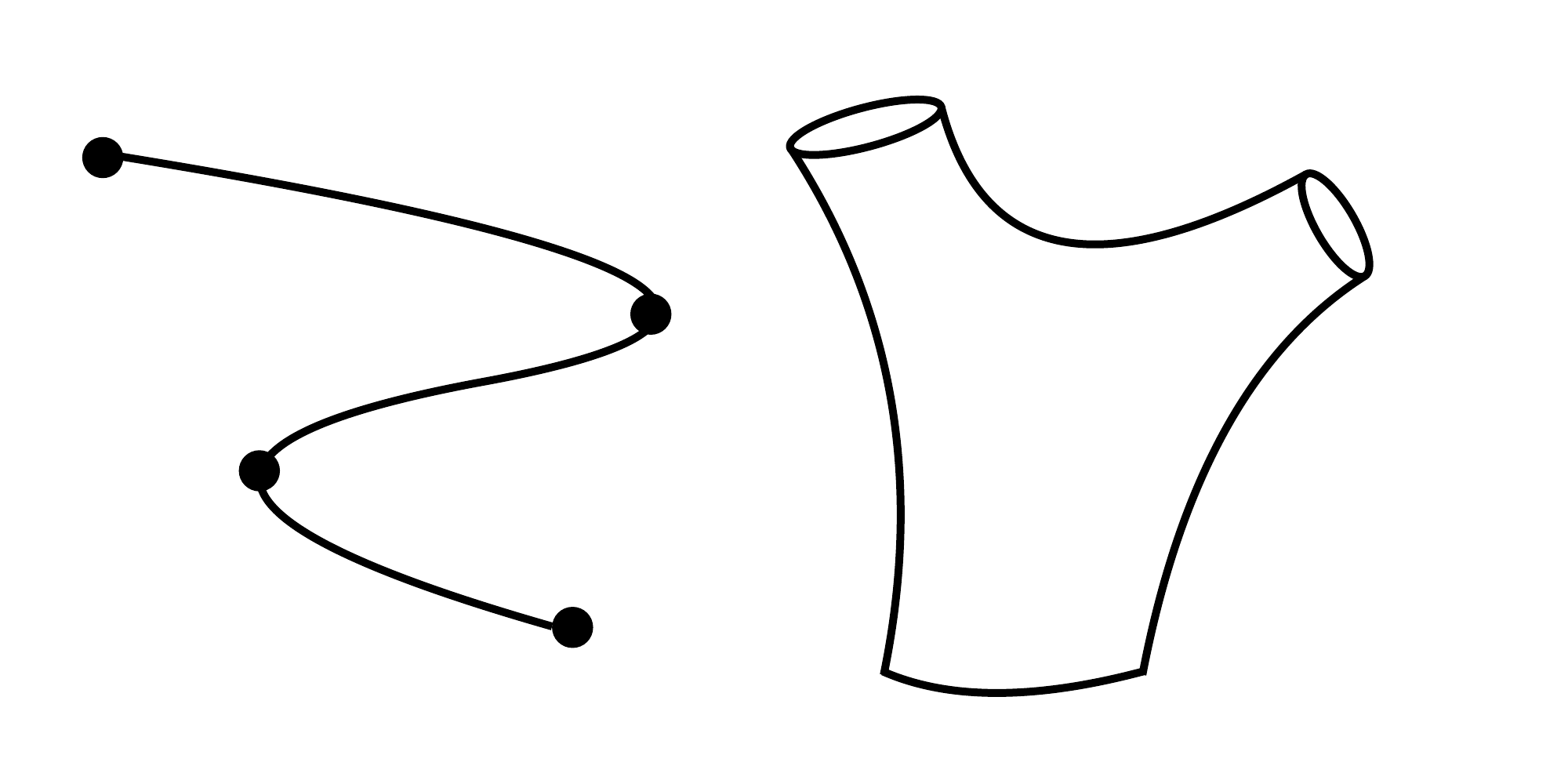}%
      \label{fig:parameterization:parameterization}
    } &
    \subfloat[Parameter Distribution]{
      \includegraphics[width=0.21\linewidth]{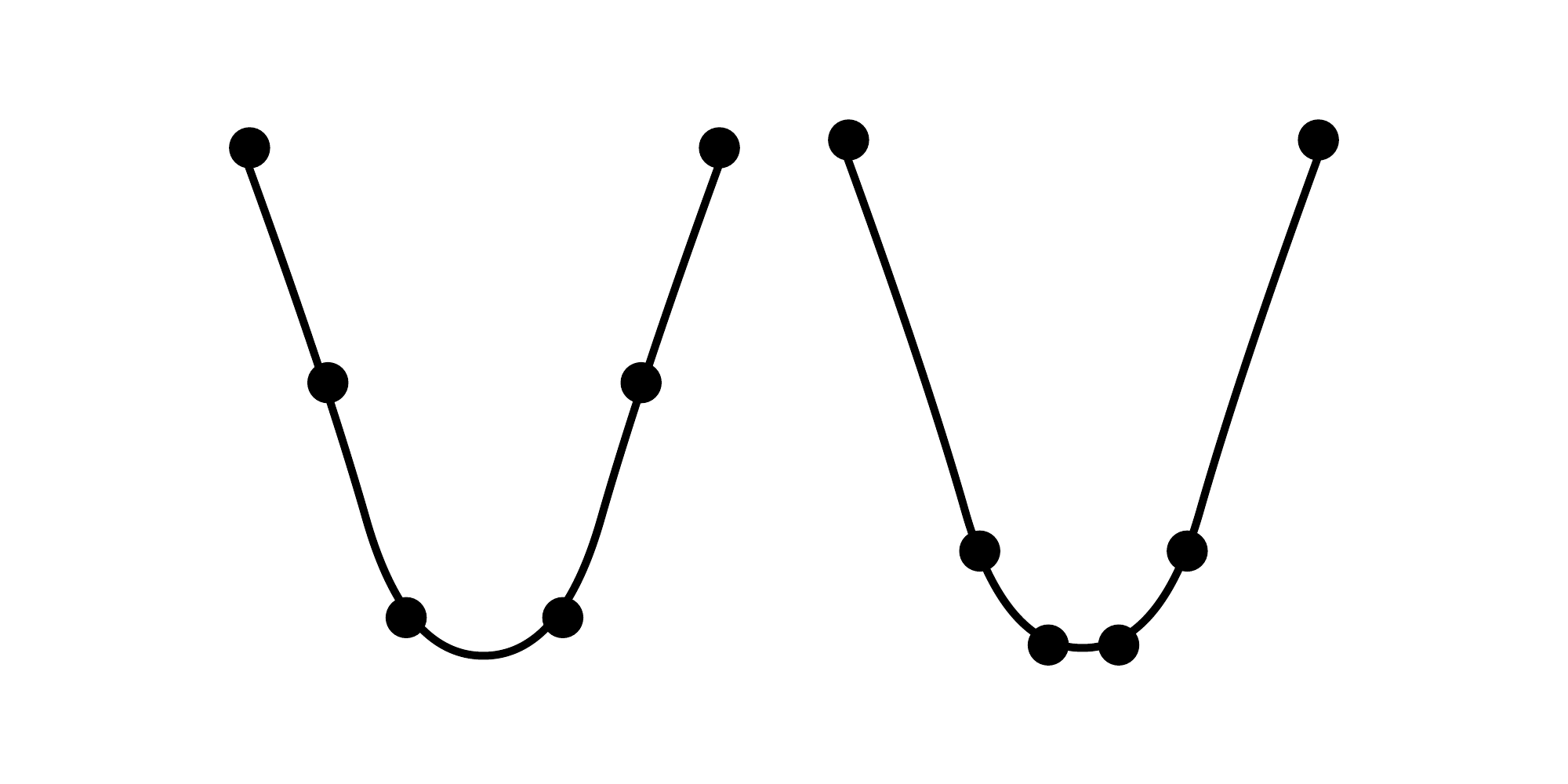}%
      \label{fig:parameterization:distribution}
    } &
    \subfloat[Topology Change]{
      \includegraphics[width=0.21\linewidth]{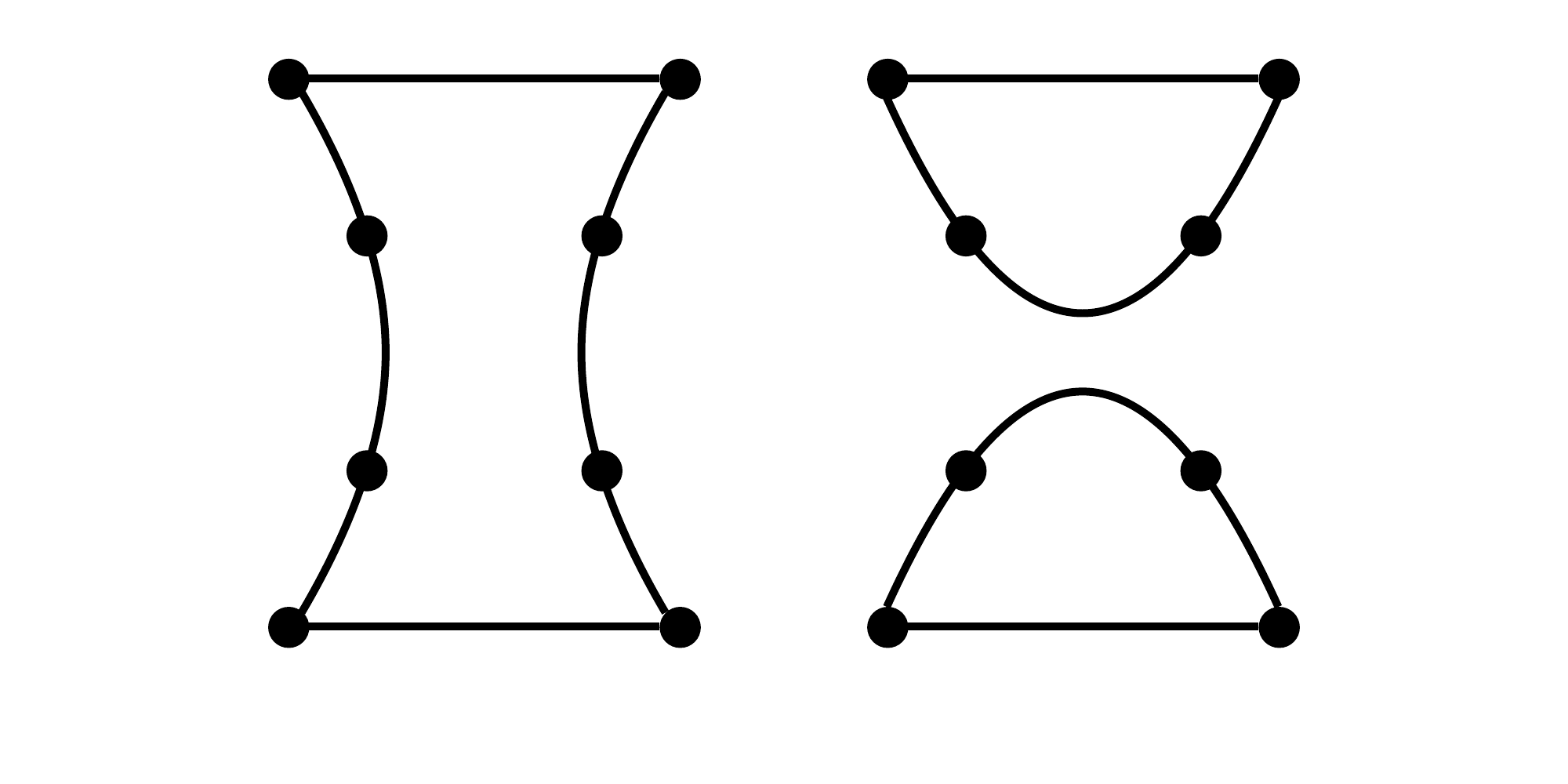}%
      \label{fig:parameterization:topology}
    } &
    \subfloat[Smooth Representation]{
      \includegraphics[width=0.21\linewidth]{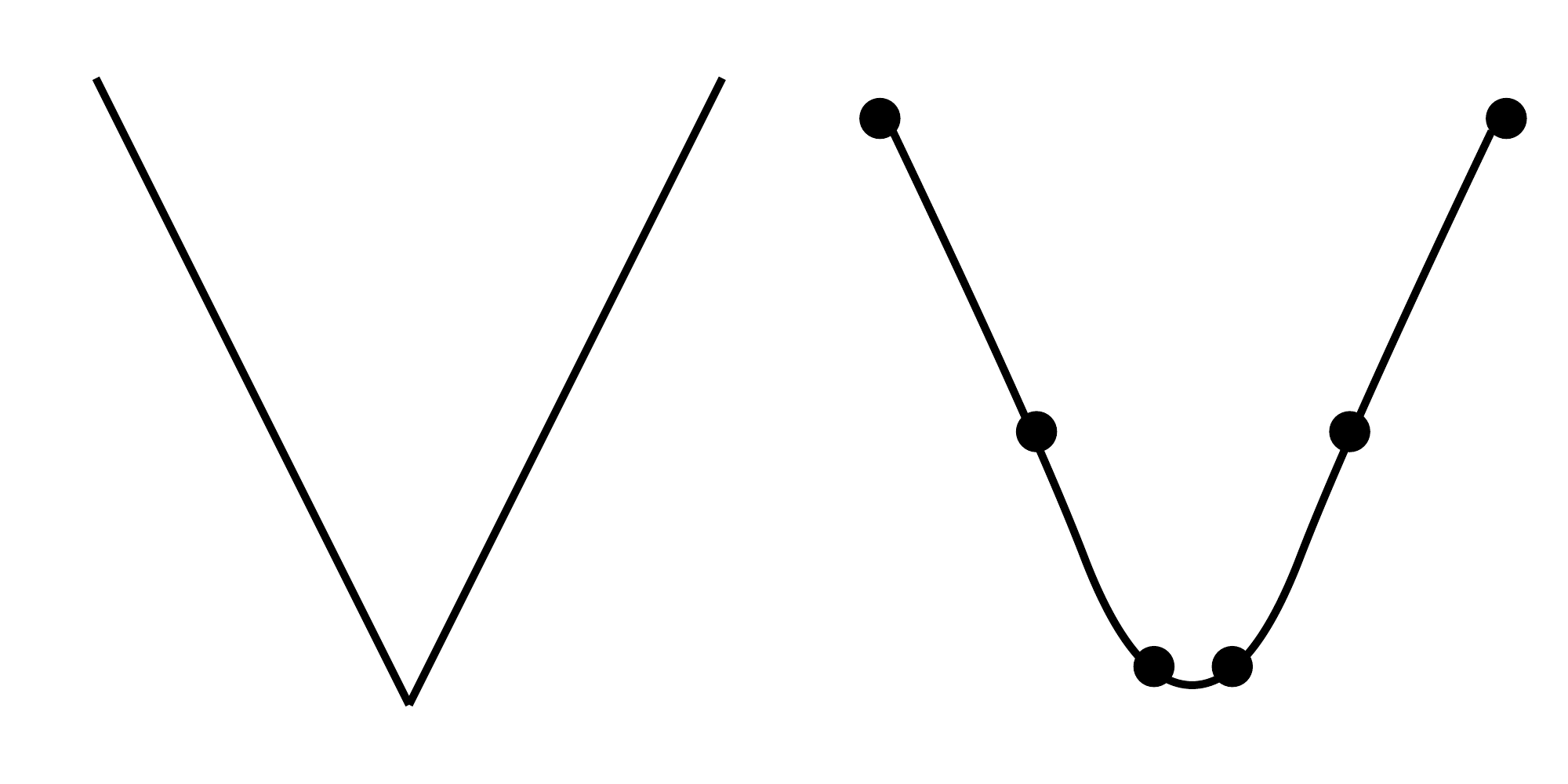}%
      \label{fig:parameterization:smooth}
    }
  \end{tabular}
  \caption{Disadvantages of using a parametric representation.~\ref{fig:parameterization:parameterization}) Complex structures such as cancellous bone are difficult to parameterize.~\ref{fig:parameterization:distribution}) Parameters will bunch during the flow.~\ref{fig:parameterization:topology}) Changes in topology require explicit merging and splitting rules.~\ref{fig:parameterization:smooth}) Surface representations are implicitely smoothed with differentiable parametric representation.}
  \label{fig:parameterization}
\end{figure*}

The two disadvantages of the proposed method are that it requires more memory and the algorithms are more difficult to implement.
The memory requirement comes from needing to store the signed distance transform in a floating point representation where binary images can be stored with an unsigned 8-bit integer and massively compressed.
Second, the proposed method requires specialized techniques for embedding and evolving, which are not yet standard in many image processing libraries.
The tools required to implement SIBA exist in virtually all image processing libraries.

Beyond contrasting, the two methods share a defining similarity: the ability to handle topological changes in bone during adaptation.
This is the defining feature of any external remodeling algorithm and follows simply as a corollary of the assumption that bone adapts at the surface.
If it adapts at the surface, topological changes will occur, and they must be treated appropriately.
Any algorithm that cannot handle topological changes, such as those assuming diffeomorphisms, are not appropriate for modeling adaptation.
This explains why successful methods in the field of brain imaging~\cite{ashburner2000voxel} and shape analysis~\cite{bookstein1997landmark} have had limited success in modeling bone adaptation.
These methods make the explicit assumption of spatial normalization: that there exists a diffeomorphism between time points.
While this appears true for brains, this is not true of bone microarchitecture within the same subject or between subjects.
Similarly, adaptation simulation methods that used deformations of the grayscale data have an implicit weak form of the topology assumption that only holds for short time durations~\cite{pauchard2008european}.

\section{Numerical Simulation}
\label{sec:numerical_simulation}

\subsection{Level Set Method}
The level set method is used to simulate the geometric flow~\cite{dervieux1980fem,dervieux1981multifluid,osher1988fronts}.
Primarily, the level set method represents the curve implicitly as the zero level set of an embedding function, $\phi$:
\begin{equation}
  \label{eqn:zero_level_set}
  \mathcal{C} = \{ x \given \phi(x) = 0 \}
\end{equation}
This prevents many issues seen in parametric representations of surfaces~\cite{kass1988snakes}, described in Figure~\ref{fig:parameterization}.
Importantly, an equation of motion can be computed for the implicit surface based on the curve evolution equation (Equation~\ref{eqn:curve_normal}) and taking the temporal derivative of the implicit contour (Equation~\ref{eqn:zero_level_set}):
\begin{equation}
  \label{eqn:level_set_pde}
  \phi_t + F \lvert \nabla \phi \rvert = 0
\end{equation}
Using this derivation, the equivalent initial value (Cauchy) problem can be stated using the implicit embedding function:
\begin{subequations}
  \begin{numcases}{}\label{eqn:ls_ivp}
    \phi_t + F \lvert \nabla \phi \rvert = 0 \\
    \phi(x,0) = \pm d(x, C)
  \end{numcases}
\end{subequations}
Instead of working with the curve directly, motion and morphometrics will be performed on the embedding, being able to recover the curve as the zero level set of the embedding at another time.

The finite difference method will be used to solve Equation~\ref{eqn:ls_ivp} numerically.
In general, the finite volume method is inappropriate for this solver as bone adaptation is not in general a conserved, hyperbolic system.
This stems from the physiology where density does not flow through the domain like a fluid leaving or entering only at the boundary, but instead changes with sources (osteoblasts) and sinks (osteoclasts) scattered throughout the domain.
Alternatively, the finite element method could be used but will in general be too computationally intensive.
This section is motivated by considering the generalized problem of Equation~\ref{eqn:ls_ivp}:
\begin{equation}
  \phi_t = L(\phi)
\end{equation}
Justification for these techniques can be found elsewhere~\cite{sethian1999level}.

\subsubsection{Spatial Gradients}
An approximation to the operator $L$ is needed.
This is a sum of the advection and mean curvature terms.
\begin{eqnarray}
  L(\phi) &=& -a + b \kappa \\
   &=& L_\text{advection}(\phi) + L_\text{mean curvature}(\phi)
\end{eqnarray}
The mean curvature term can simply be expanded and appropriate finite difference subsituted into the equation:
\begin{eqnarray}
  L_\text{mean curvature}(\phi) = b \kappa \lvert \nabla \phi \rvert \\
   = \frac{
    \begin{aligned}
      \left(\phi_{yy} + \phi_{zz}\right)\phi_x^2 \\
    + \left(\phi_{zz} + \phi_{xx}\right)\phi_y^2 \\
    + \left(\phi_{xx} + \phi_{yy}\right)\phi_z^2 \\
    -2 \phi_x\phi_y\phi_{xy}
    -2 \phi_z\phi_x\phi_{zx}
    -2 \phi_y\phi_z\phi_{yz}
  \end{aligned}
  }{
  \left(\phi_x^2 + \phi_y^2 + \phi_z^2\right)
  }
\end{eqnarray}
The advection term is more difficult because central-difference approximations to the gradient operator cause oscillations.
Instead, an upwind solver must be used:
\begin{eqnarray}
  L_\text{advection}(\phi) = - a \lvert \nabla \phi \rvert \\
  = -a \sqrt{\phi_x^2 + \phi_y^2 + \phi_z^2} \\
  \begin{split}
  = - a^+ \sqrt{(p^+)^2 + (q^-)^2 + (r^+)^2 + (s^-)^2 + (t^+)^2 + (u^-)^2}\\
  - a^- \sqrt{(p^-)^2 + (q^+)^2 + (r^-)^2 + (s^+)^2 + (t^-)^2 + (u^+)^2}
  \end{split}
\end{eqnarray}
where $x^+ = \max(x,0)$, $x^- = \min(x,0)$, and $p$ through $u$ are one-sided differences:
\begin{equation}
  \begin{matrix}
    p = D_x^-\phi & q = D_x^+\phi \\
    r = D_y^-\phi & s = D_y^+\phi \\
    t = D_z^-\phi & u = D_z^+\phi 
  \end{matrix}
\end{equation}
These first-order derivatives can be replaced with weighted essential non-oscillator schemes if higher order accuracy is needed~\cite{liu1994weighted}.

\subsubsection{Time Stepping}
Next, the solution must be time stepped.
This is done using a forward Euler approximation to the time derivative:
\begin{equation}
  \phi^{n+1} = \phi^n + \Delta t L(\phi^n)
\end{equation}
This can be extended with the Runge-Kutta method if a more accurate solver is needed.

\subsubsection{Courant-Friedrich-Lewy Condition}
The method will be unstable if the Courant-Friedrich-Lewy (CFL) condition is not met~\cite{courant1928partiellen}.
The CFL condition states that the ``numerical domain of dependence must include the physical domain of dependence''.
In essence, this means that the surface cannot travel further than a voxel in a single iteration.
The time step can be selected by the following equation:
\begin{equation}
  \alpha = 
  \Delta t \left(\frac{\lvert a \rvert}{\min(\Delta x, \Delta y, \Delta z)} + \frac{2 \lvert b \rvert}{\min(\Delta x^2, \Delta y^2, \Delta z^2)}\right)
\end{equation}
where $\Delta x$, $\Delta y$, and $\Delta z$ are the voxels edge lengths.
$\alpha$ must be selected less than $1$ to satisfy the CFL condition and is selected to be $0.5$ in this work.

\subsubsection{Reinitialization}
\label{subsubsec:reinitialization}
During evolution, the embedding can deviate from a signed distance transform.
Reinitialization is the process of returning the embedding to a signed distance transform~\cite{sussman1994level}.
In this work, a method is used that guarantees that the embedding does not change sign during reinitalization, and thus keeps volume conserved~\cite{peng1999pde}.
This is achieved by solving the following partial differential equation with the same finite difference method:
\begin{equation}
  \phi_\tau + S(\phi_0) \left(\lvert \nabla \phi \rvert - 1\right) = 0
\end{equation}
where $\tau$ is a time-like parameter and $S(\cdot)$ is a regularized approximation to the sign function:
\begin{equation}
  S(\phi) = \frac{\phi}{\sqrt{\phi^2 + \lvert \nabla \phi \rvert^2 h^2}}
\end{equation}
where $h = \min(\Delta x, \Delta y, \Delta z)$.
The reinitalization equation is performed after every iteration, which could be relaxed if computation time was a concern.

\subsection{Embedding}
Attention is now placed on initialization the embedding, $\phi$.
This is a challenging task as the signed distance transform of binary images exhibit quantization, limiting numerical accuracy of the flow and ability to measure curvatures~\cite{besler2020artifacts}.
Instead, a previously developed method is used that instantiates the embedding directly from the density image, skipping binarization~\cite{besler2021constructing}.
The method is quickly summarized below.

The central idea is to construct an embedding $\psi$ that does not satisfy the Eikonal condition but shares a zero crossing with the desired embedding $\phi$.
This is achieved by subtracting a density threshold to shift the desired density level set to zero.
Furthermore, noise reduction methods can also be applied.
This leads to the following definition of the intermediate embedding:
\begin{equation}
  \psi = T - G_\sigma * \rho
\end{equation}
where $T$ is a density threshold, $G_\sigma$ is a Gaussian blur of size $\sigma$, and $\psi$ is an embedding whose zero level set is the implicit surface (Equation~\ref{eqn:zero_level_set}).
Having the intermediate embedding, the closest point method~\cite{chopp2001some,ruuth2008simple,coquerelle2016fourth} is used to establish the narrowband distances, which are then marched to the remainder of the domain using the high-order fast sweeping method~\cite{zhang2006high}.
The obtained embedding $\phi$ satisfies the recovery condition, Eikonal condition, is unique, and has an order of accuracy greater than unity~\cite{besler2021highorder,besler2021constructing}.

\subsection{Density Component Estimation}
Having the density and embedding image, the phase densities $\rho_\text{marrow}$ and $\rho_\text{bone}$ can be estimated.
The central idea is that the density image can be constructed from the embedding image knowing the two phase densities and the embedding:
\begin{equation}
  \rho = \rho_\text{bone} \theta(-\phi) + \rho_\text{marrow} \left[1 - \theta(-\phi)\right]
\end{equation}
where $\theta$ is the Heaviside step function.
We follow here the method of Chan and Vese~\cite{chan2001active} to estimated $\rho_\text{marrow}$ and $\rho_\text{bone}$ from the density image $\rho$ an embedding $\phi$:
\begin{eqnarray}
  \label{eqn:rho_bone}
  \rho_\text{bone} &=& \frac{\int_\Omega \rho(x) \theta(-\phi(x)) dV}{\int_\Omega \theta(-\phi(x)) dV} \\
  \label{eqn:rho_marrow}
  \rho_\text{marrow} &=& \frac{\int_\Omega \rho(x) \left[1 - \theta(-\phi(x))\right] dV}{\int_\Omega \left[1 - \theta(-\phi(x))\right] dV}
\end{eqnarray}
It is assumed that the phase densities do not change during adaptation.
As such, volumetric bone mineral density can be estimated from the volume fraction of bone:
\begin{eqnarray}
  \text{BV/TV} &=& \frac{\int_\Omega \theta(-\phi(x)) dV}{\int_\Omega dV} \\
  \text{vBMD} &=& \rho_\text{bone} \text{BV/TV} + \rho_\text{marrow} \left[1 - \text{BV/TV}\right]
\end{eqnarray}
We remark here that while masking is not explicitly performed in this study, it can be achieved using simple surface editing operators on signed distance fields~\cite{zhao2000implicit,museth2002level}.

\subsection{Morphometry}
Beyond densities, morphometry can be performed directly from the embedding using a previously developed technique~\cite{besler2021morph}.
First, mean curvature can be computed using the divergence of the surface normals:
\begin{equation}
  H = \frac{1}{2} \nabla \cdot \left( \frac{\nabla \phi}{\lvert \nabla \phi \rvert} \right)
\end{equation}
We remark that $H$ and $\kappa$ differ by a factor of the surface dimensionality:
\begin{equation}
  H = \frac{\kappa}{2}
\end{equation} 
$\kappa$ is the physicist's mean curvature while $H$ is the geometer's mean curvature.
Next, Gaussian curvature can be computed on the implicit contour~\cite{sethian1999level}:
\begin{equation}
  K = -\frac{\begin{vmatrix}
    \phi_{xx} & \phi_{xy} & \phi_{xz} & \phi_x \\
    \phi_{yx} & \phi_{yy} & \phi_{yz} & \phi_y \\
    \phi_{zx} & \phi_{zy} & \phi_{zz} & \phi_z \\
    \phi_x & \phi_y & \phi_z & 0
    \end{vmatrix}}{\lvert \nabla \phi \rvert^4}
\end{equation}
Using mean and Gaussian curvature with definition of the volume and area elements, the volume ($V$), area ($A$), surface average mean curvature ($\langle H \rangle$), total Gaussian curvature ($\bar{K}$), and \epc~characteristic ($\chi$) can be measured using previously developed methods~\cite{chan2001active,peng1999pde,besler2021morph,besler2021constructing}:
\begin{eqnarray}
  V &=& \int_\Omega \theta(-\phi) dV \\
  A &=& \int_\Omega \delta(-\phi) |\nabla \phi| dV \\
  \langle H \rangle &=& \frac{\int_\Omega H \delta(-\phi) |\nabla \phi| dV}{\int_\Omega dV} \\
  \bar{K} &=& \int_\Omega K \delta(-\phi) |\nabla \phi| dV \\
  \chi &=& \frac{\bar{K}}{2\pi}
\end{eqnarray}

From these definitions, standard bone morphometric measures can be derived.
The structure model index (SMI) can be computed from average mean curvature, volume, and area~\cite{hildebrand1997quantification,jinnai2002surface}, trabecular bone pattern factor (TBPf) can be computed from average mean curvature alone~\cite{hahn1992trabecular,stauber2006volumetric}, and connectivity density (Conn.D) can be computed from the \epc~characteristic and the total volume~\cite{odgaard1993quantification}:
\begin{eqnarray}
  \text{SMI} &=& 12 \langle H \rangle \frac{V}{A} \\
  \text{TBPf} &=& 2 \langle H \rangle \\
  \text{Conn.D} &=& \frac{1 - \chi/2}{TV}
\end{eqnarray}
The purpose of the factor-of-two is outlined in the Appendix.
Importantly, since connected component filtering cannot be easily implemented on the embedding and disconnected particles will form during the flow, Odgaard's constraints on the Betti numbers --- that there is no marrow cavities ($\beta_2 = 0$) and only one foreground particle ($\beta_0 = 1$) --- cannot be guaranteed~\cite{odgaard1993quantification}.
The importance of this conclusion is outlined in the discussion.

\section{Experiment}
Two experiments are performed to demonstrate curvature-based bone adaptation.
Ten cubes of bovine trabecular bone were previously sawed to 10 mm in edge length and imaged at a nominal resolution of $\SI{20}{\micro\metre}$~\cite{sandino2013trabecular}.
Embedding was performed with a threshold of $T = 400~\text{mg HA/cc}$ and Gaussian filter of standard deviation $\sigma = \SI{20}{\micro\metre}$.
This dataset was previously used to validate the embedding method~\cite{besler2021constructing}.

Images were embedded and geometric flows simulated using the level set method.
Two parameter sets are studied as described later.
30 years were simulated and morphometry was performed every 3 years directly from the embedding.
Morphometry included the bone surface to volume ratio (BS/BV, $\si{\per\milli\metre}$), volumetric bone mineral density (vBMD, $\text{mg HA/cc}$), connectivity density (Conn.D, $\si{\per\milli\metre\cubed}$), and structure model index (SMI, $-$).
For rendering, volumes were reduced to a $\SI{2}{\milli\metre}$ edge length cube in order to visualize individual trabeculae and the marching cubes algorithm~\cite{lorensen1987marching} was used to directly extract the surface at an isocontour of zero.

\subsection{Assigned Flow}
Model parameters are selected such that they represent physiologically plausible losses.
Selecting $a = \SI{-1}{\micro\metre\per\year}$ and $b = \SI{100}{\micro\metre\squared\per\year}$ would erode a rod of thickness $\SI{100}{\micro\metre}$ roughly $\SI{2}{\micro\metre\per\year}$.
Given that trabecular bone has an average thickness around $100 - 250~\si{\micro\metre}$, this is a reasonable loss over a human lifespan of 100 years.
Of course, the process is non-linear, and more loss will be experienced.

\begin{table}
  \centering
  \begin{tabular}{ccccc}
    \hline
    Subject & $\langle H \rangle~\si{\per\milli\metre}$ & $\kappa~\si{\per\milli\metre}$ & $b~\si{\micro\metre\squared\per\year}$	& $a~\si{\micro\metre\per\year}$ \\
    \hline
    1 & 2.21 & 4.42 & 100 & 0.442 \\
    2 & -0.13 & -0.26 & 100 & -0.026 \\
    3 & 1.27 & 2.53 & 100 & 0.253 \\
    4 & 0.31 & 0.62 & 100 & 0.062 \\
    5 & 0.37 & 0.73 & 100 & 0.073 \\
    6 & 1.25 & 2.51 & 100 & 0.251 \\
    7 & 0.56 & 1.12 & 100 & 0.112 \\
    8 & 2.11 & 4.22 & 100 & 0.422 \\
    9 & 1.35 & 2.70 & 100 & 0.270 \\
    10 & -0.02 & -0.04 & 100 & -0.004 \\
   \hline
  \end{tabular}
  \caption{Per-subject parameters for curvature driven bone adaptation that produce a volume-preserving flow.}
  \label{table:parameters}
\end{table}

\subsection{Volume-Preserving Flow}
Next, a volume-preserving flow is investigated.
To ensure the flow is volume-preserving, the parameters $\left\{a, b\right\}$ are selected equal to the average mean curvature across the surface:
\begin{equation}
  \frac{a}{b} = \langle \kappa \rangle
\end{equation}
The same curvature constant, $b = \SI{100}{\micro\metre\squared\per\year}$, is used but the propagation constant, $a$, is allowed to vary with each subject.
These values are given in Table~\ref{table:parameters}.
The volume-preserving flow is interesting because it implies changes in the bone surface that fundamentally cannot be measured by areal or volumetric bone volume fraction, requiring imaging of the microarchitecture.
Furthermore, this suggests a deeper relationship between structure and calcium homeostasis where bone can be turning over but net calcium flux through digestion and excretion is zero.

\begin{figure*}
  \centering
  \begin{tabular}{cc}
    \subfloat[BS/BV]{
      \includegraphics[width=0.45\linewidth]{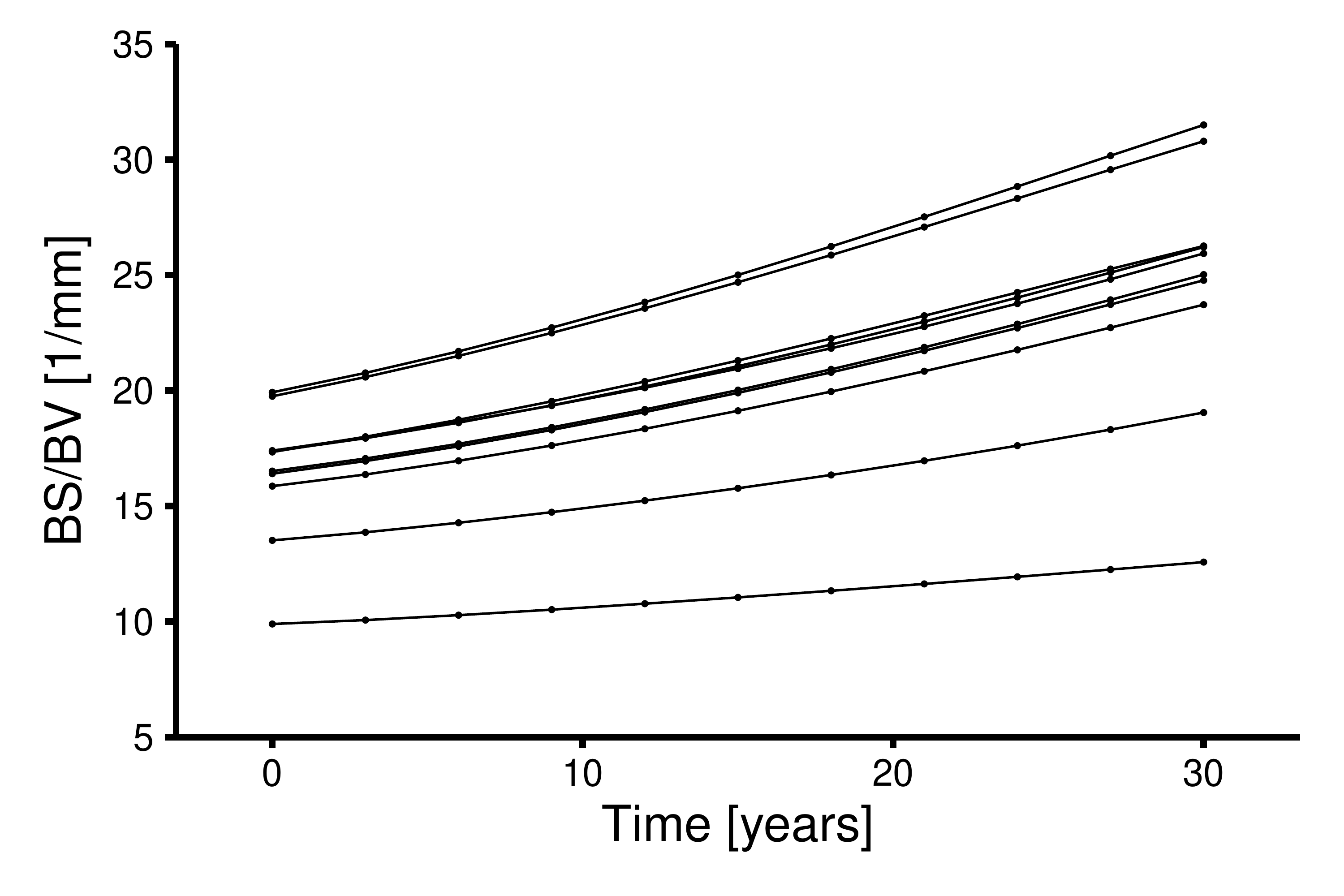}%
      \label{fig:flow:bsbv}
    } &
    \subfloat[vBMD]{
      \includegraphics[width=0.45\linewidth]{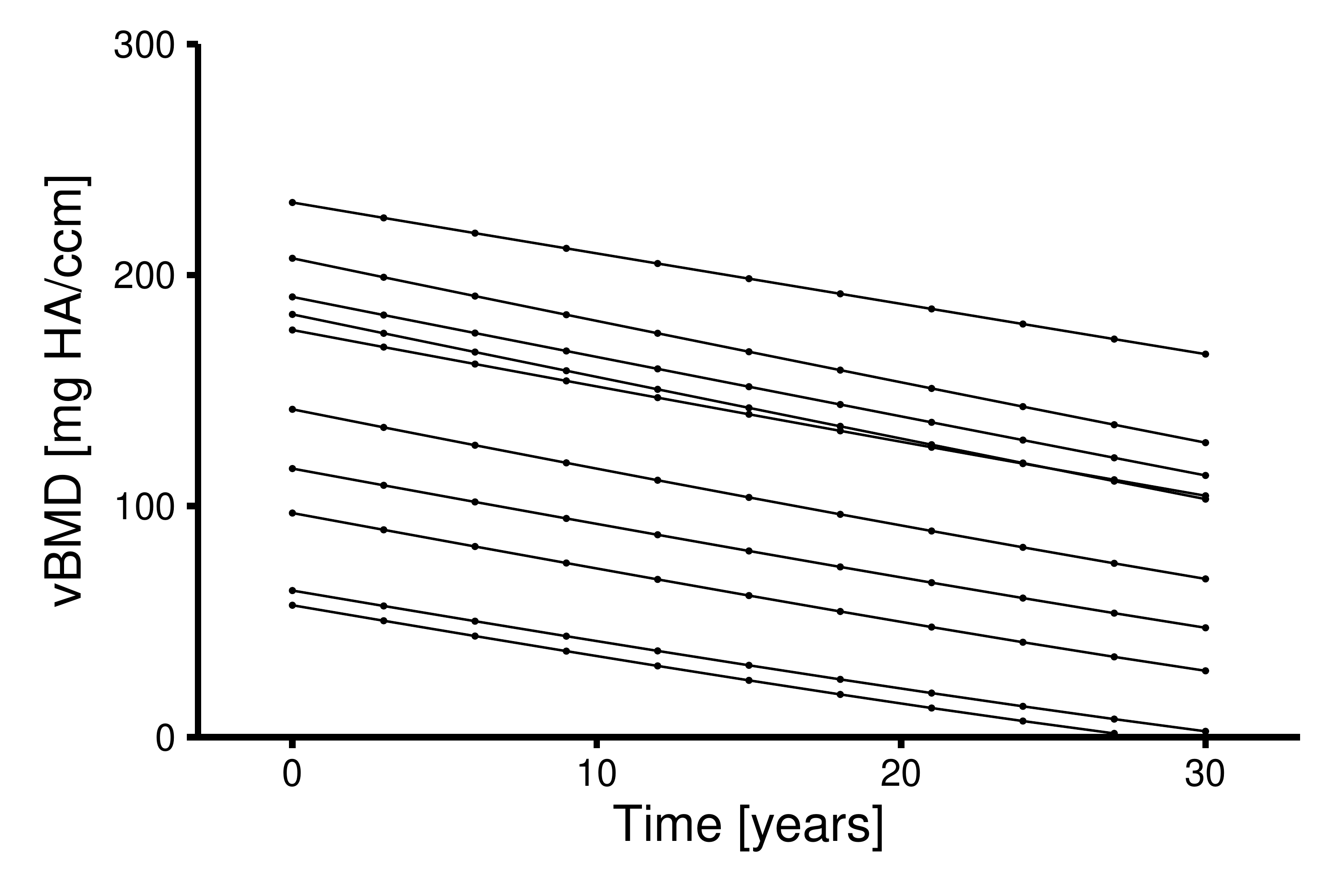}%
      \label{fig:flow:vbmd}
    } \\
    \subfloat[Conn.D]{
      \includegraphics[width=0.45\linewidth]{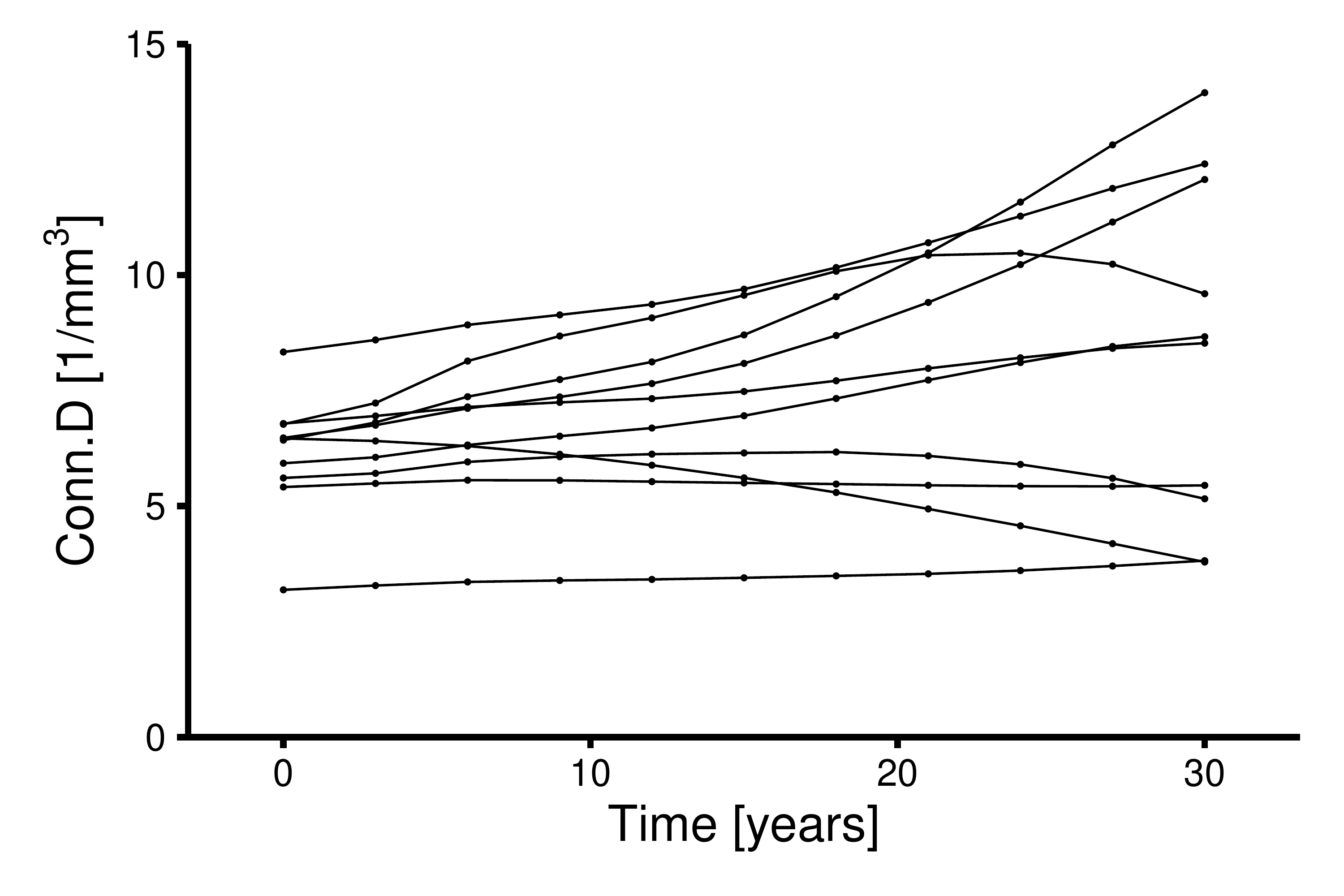}%
      \label{fig:flow:connd}
    } &
    \subfloat[SMI]{
      \includegraphics[width=0.45\linewidth]{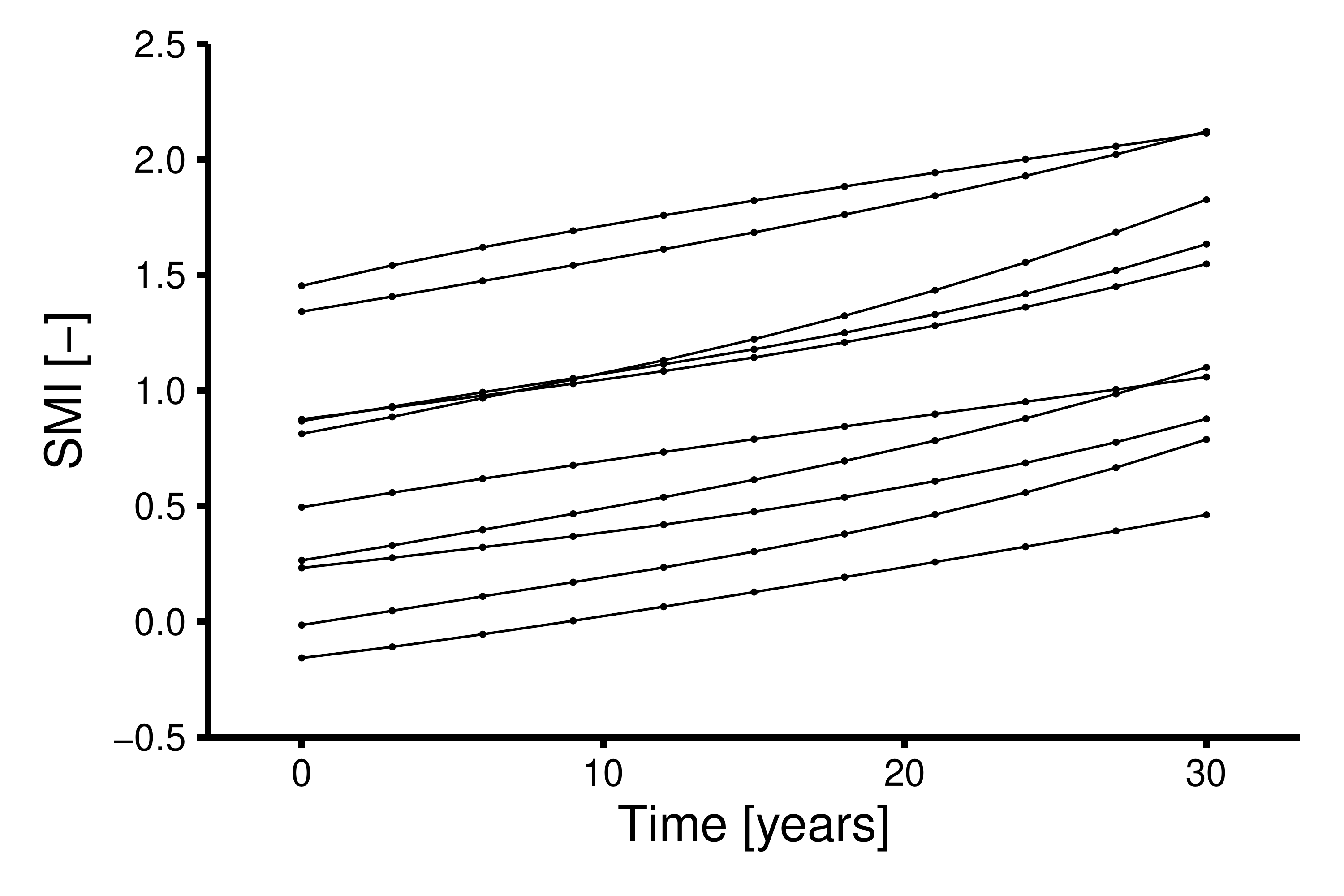}%
      \label{fig:flow:smi}
    }
  \end{tabular}
  \caption{Measured morphometry during curvature based bone adaptation with parametes $a = \SI{-1}{\micro\metre\per\year}$ and $b = \SI{100}{\micro\metre\squared\per\year}$. Bone (\ref{fig:flow:bsbv}) surface area to volume ratio, (\ref{fig:flow:vbmd}) volumetric minearl density, (\ref{fig:flow:connd}) connectivity density, and (\ref{fig:flow:smi}) structure model index are plotted every three years over 30 years of simulation. Connected lines are individual subjects.}
  \label{fig:flow}
\end{figure*}

\begin{figure*}[h!]
  \centering
  \begin{tabular}{cccc}
    \subfloat[$t = \SI{0}{\year}$]{
      \includegraphics[width=0.15\linewidth]{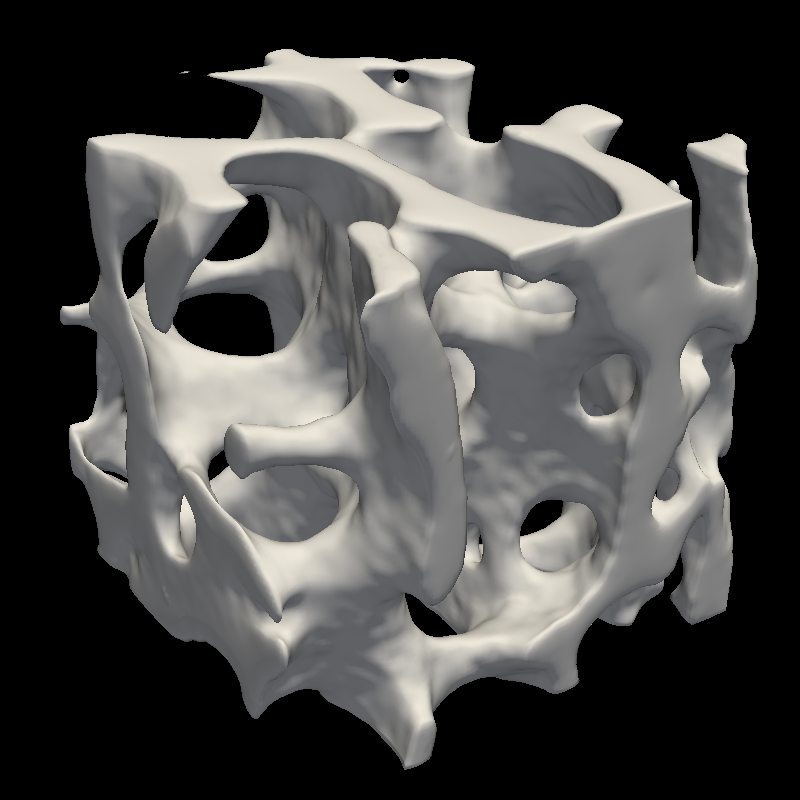}%
      \label{fig:times:0}
    } &
    \subfloat[$t = \SI{3}{\year}$]{
      \includegraphics[width=0.15\linewidth]{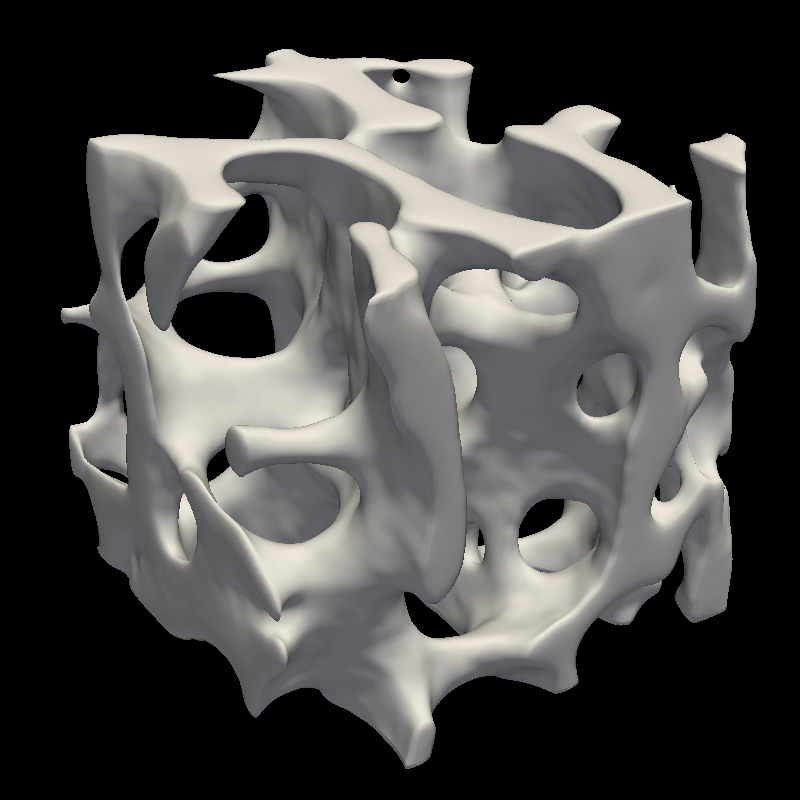}%
      \label{fig:times:3}
    } &
    \subfloat[$t = \SI{6}{\year}$]{
      \includegraphics[width=0.15\linewidth]{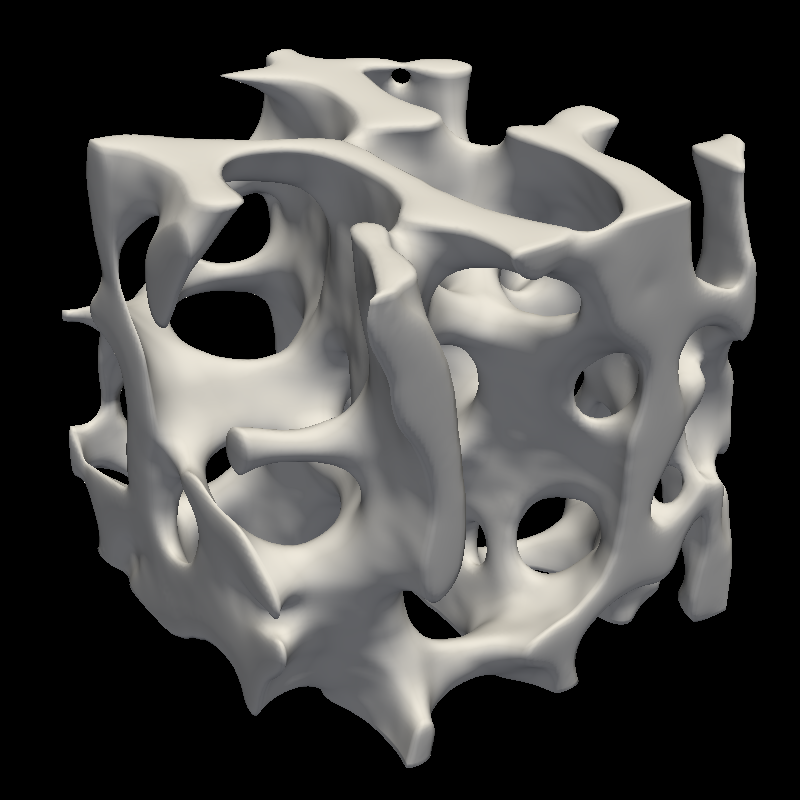}%
      \label{fig:times:6}
    } &
    \subfloat[$t = \SI{9}{\year}$]{
      \includegraphics[width=0.15\linewidth]{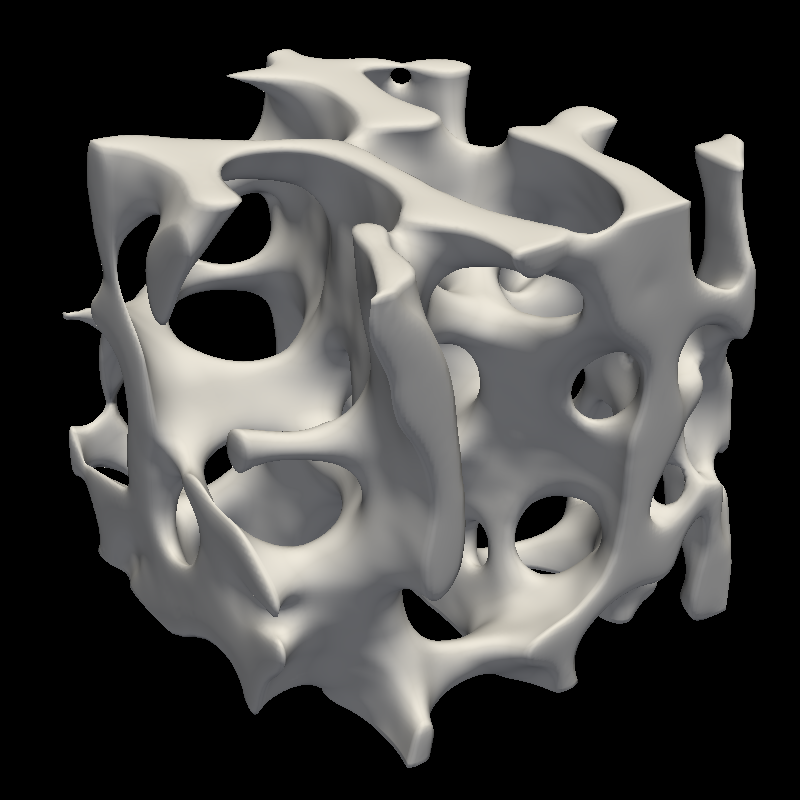}%
      \label{fig:times:9}
    } \\
    \subfloat[$t = \SI{12}{\year}$]{
      \includegraphics[width=0.15\linewidth]{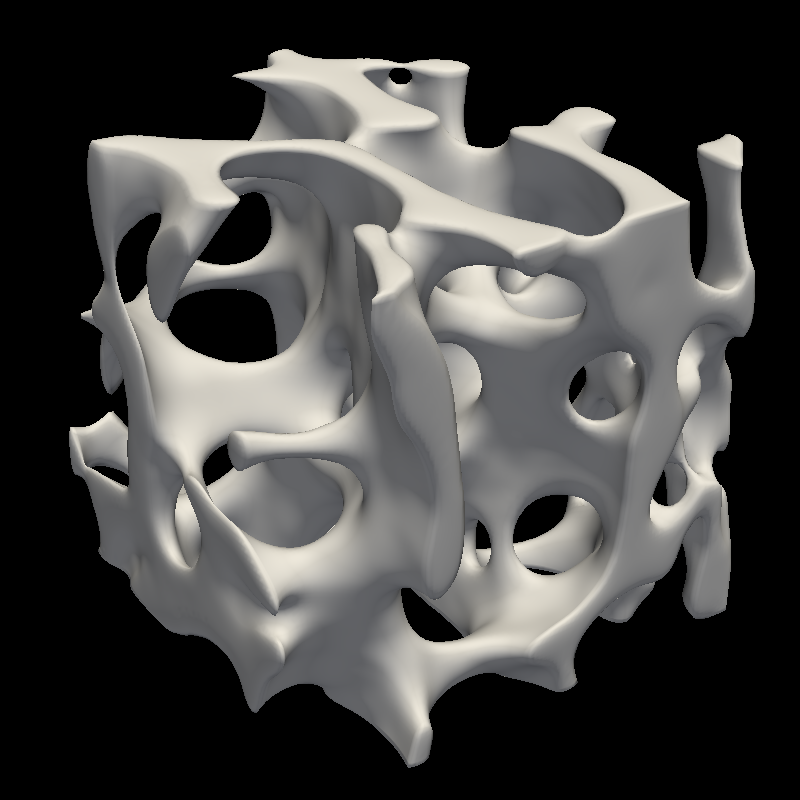}%
      \label{fig:times:12}
    } &
    \subfloat[$t = \SI{15}{\year}$]{
      \includegraphics[width=0.15\linewidth]{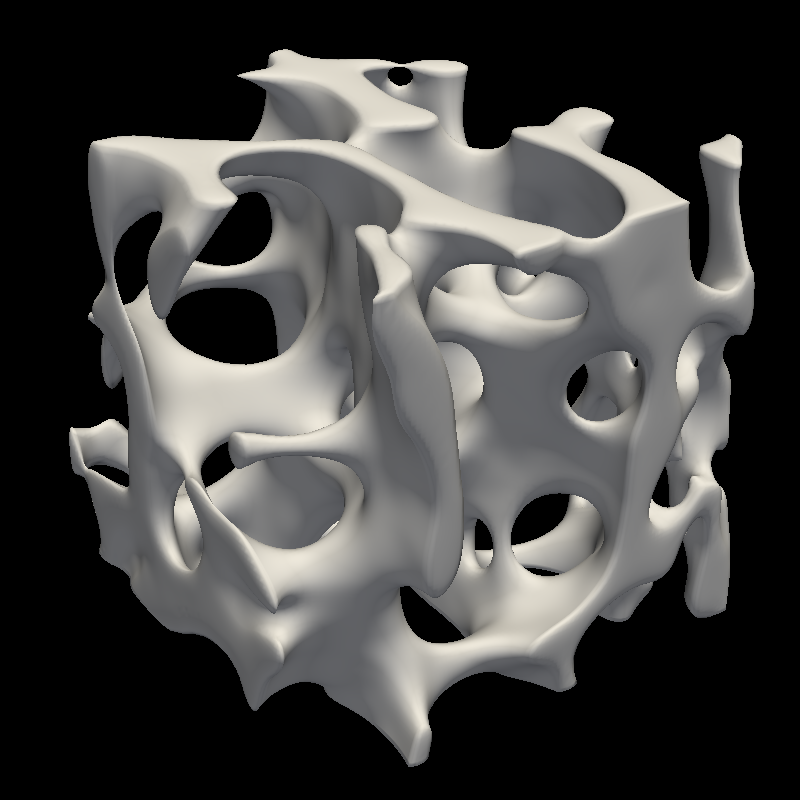}%
      \label{fig:times:15}
    } &
    \subfloat[$t = \SI{18}{\year}$]{
      \includegraphics[width=0.15\linewidth]{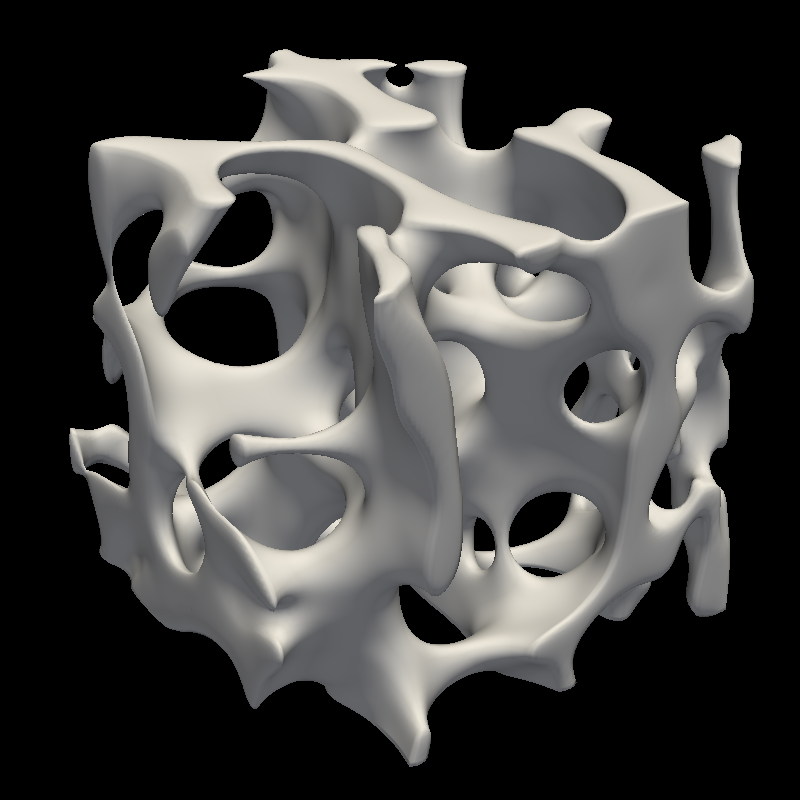}%
      \label{fig:times:18}
    } &
    \subfloat[$t = \SI{21}{\year}$]{
      \includegraphics[width=0.15\linewidth]{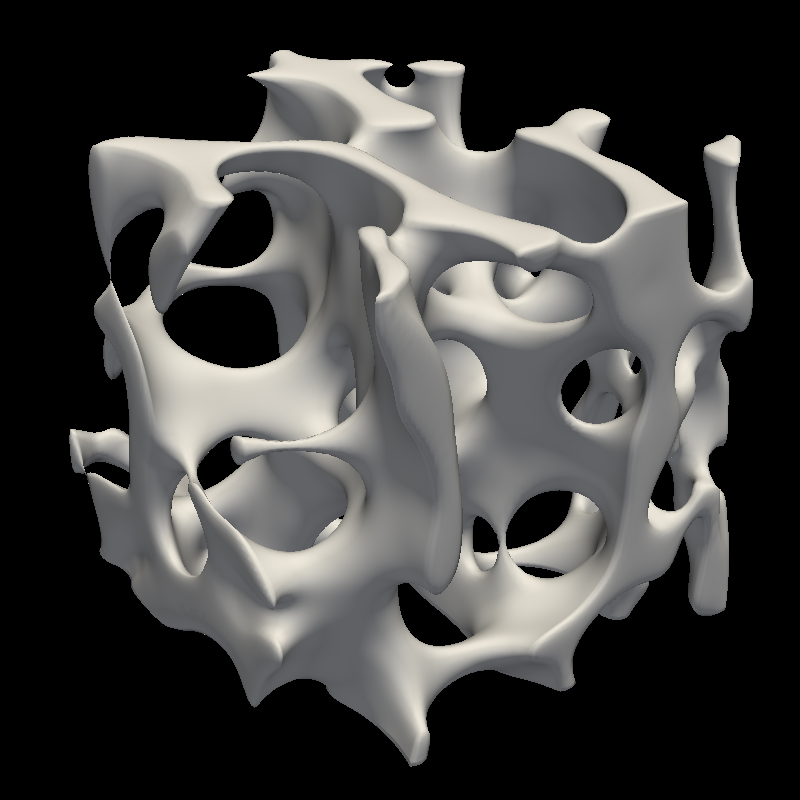}%
      \label{fig:times:21}
    } \\
    \subfloat[$t = \SI{24}{\year}$]{
      \includegraphics[width=0.15\linewidth]{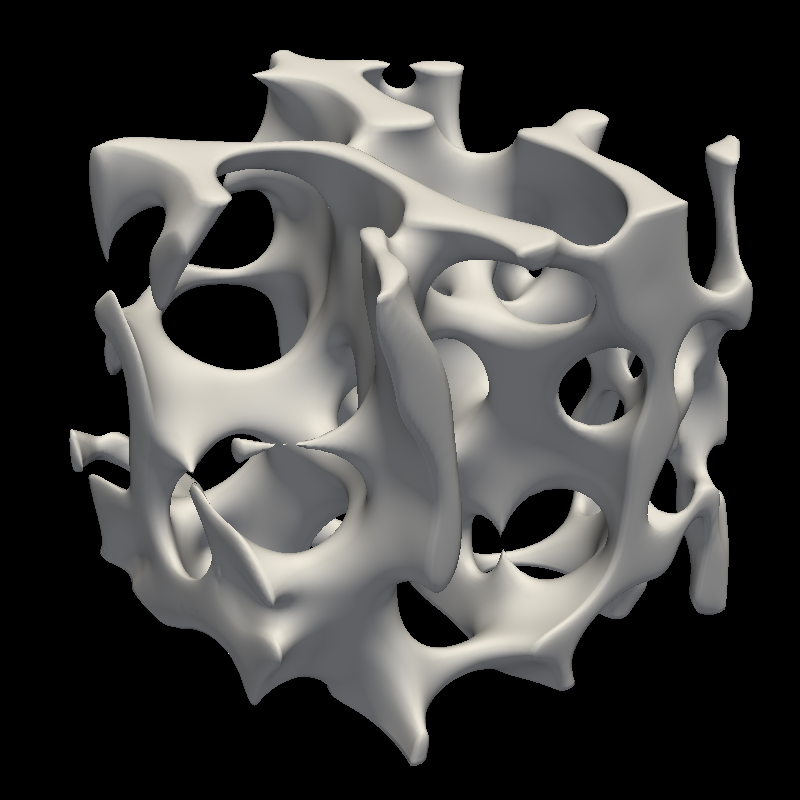}%
      \label{fig:times:24}
    } &
    \subfloat[$t = \SI{27}{\year}$]{
      \includegraphics[width=0.15\linewidth]{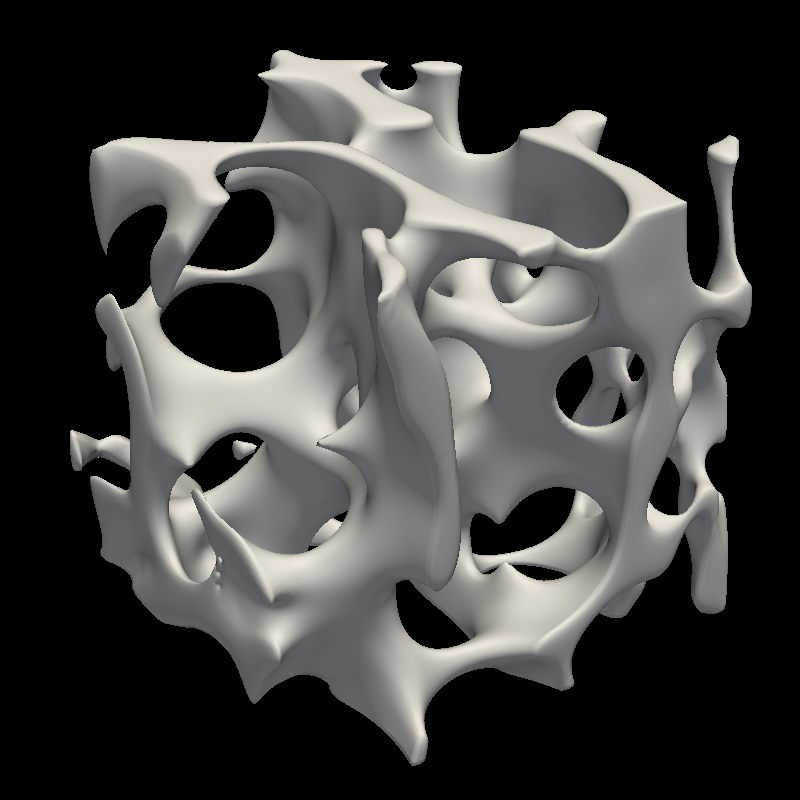}%
      \label{fig:times:27}
    } &
    \subfloat[$t = \SI{30}{\year}$]{
      \includegraphics[width=0.15\linewidth]{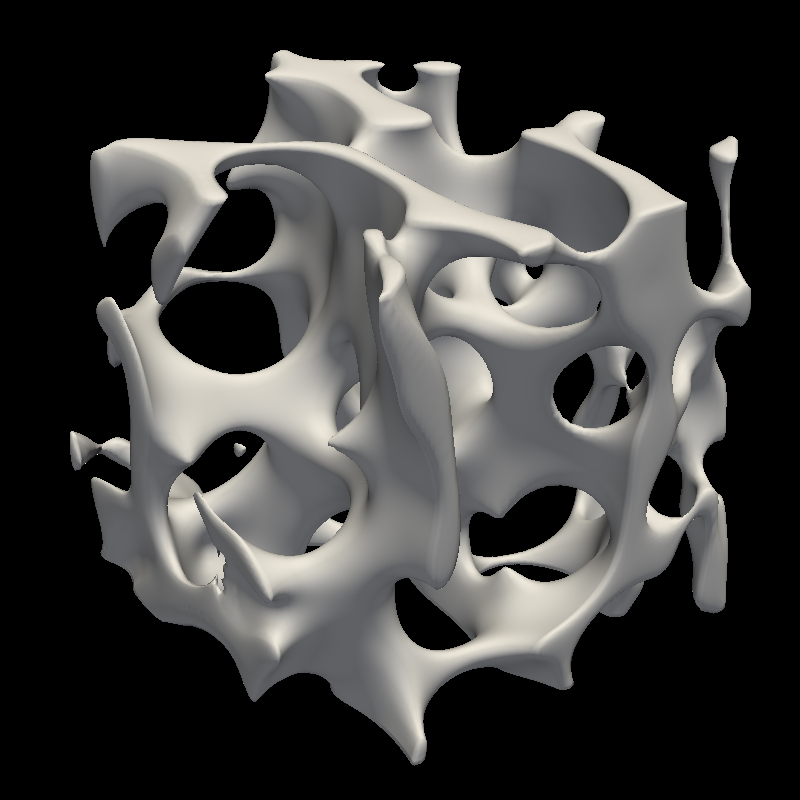}%
      \label{fig:times:30}
    } &
  \end{tabular}
  \caption{Visualization of the bone surface changing across the simulation timeframe. Rods disconnect exhibiting the ability of level set methods to capture topological changes. The medial subject by bone mineral density is displayed.}
  \label{fig:times}
\end{figure*}

\section{Results}

\subsection{Assigned Flow}
Morphometry during curvature-based bone adaptation is plotted in Figure~\ref{fig:flow}.
Bone surface to volume ratio increases non-linearly with time.
Similarly, structure model index increases with time.
Owing to the inverse relationship between SMI and BS/BV, the changes in SMI must be driven by an increase in mean curvature across the surface.
Bone mineral density decreases almost linearly and nearly at the same rate for all subjects.
A subject specific response was seen in connectivity density.
While a few subjects increased Conn.D across the time frame, others increased then decreased rapidly.
The number of connections can rise when plates form holes, decrease when rods disconnect, and effectively decrease by the formation of isolated particles, making this morphometric outcome difficult to interpret.
The surface of the median subject by density is visualized in Figure~\ref{fig:times}.
Rods are seen disconnecting and resorbing, the structure thins throughout and becomes more porous.
Furthermore, the surface looses roughness in the first 3 years consistent with noise being removed by mean curvature flow.

\subsection{Volume-Preserving Flow}
\begin{figure*}[t]
  \centering
  \begin{tabular}{cc}
    \subfloat[BS/BV]{
      \includegraphics[width=0.45\linewidth]{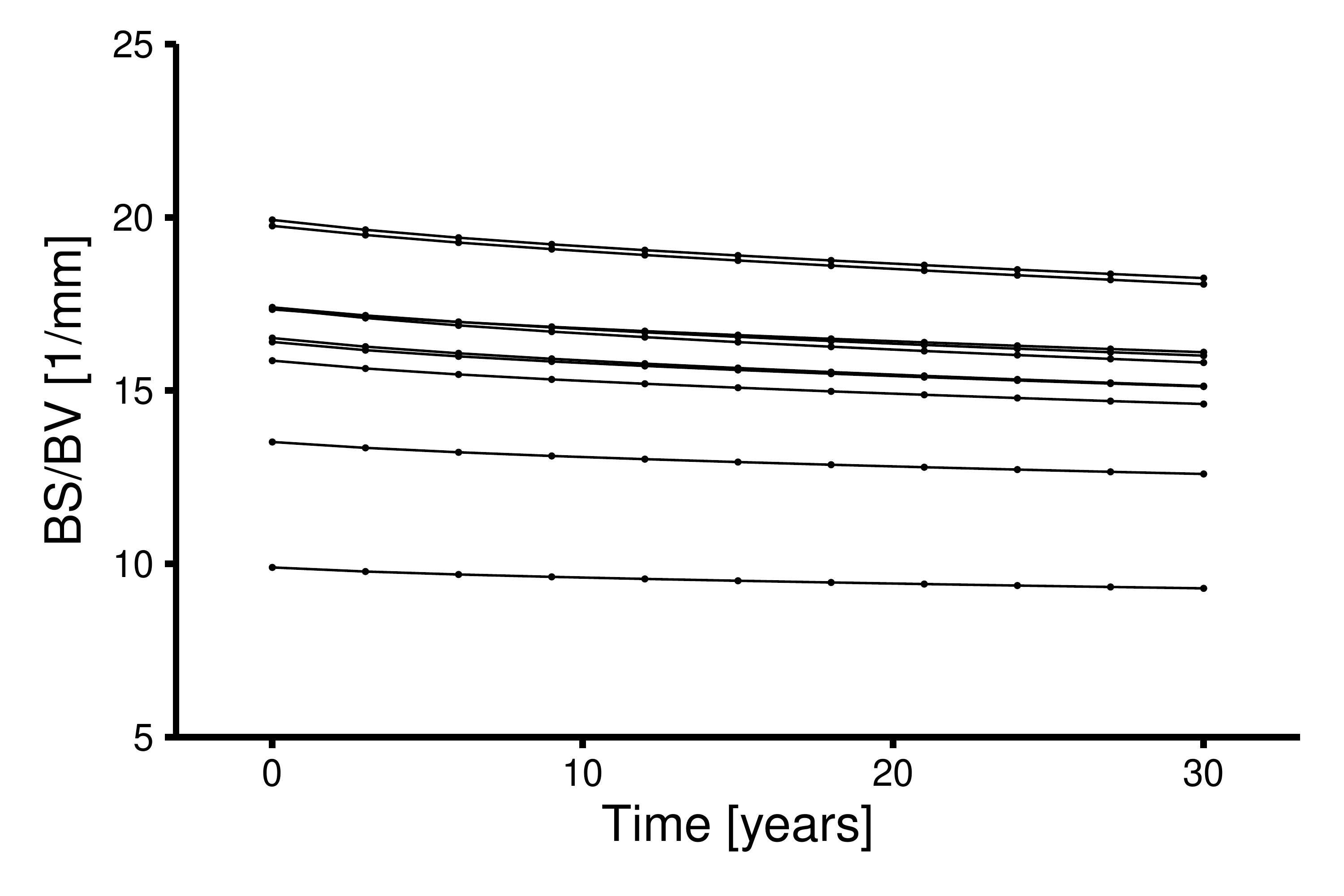}%
      \label{fig:morph:bsbv}
    } &
    \subfloat[vBMD]{
      \includegraphics[width=0.45\linewidth]{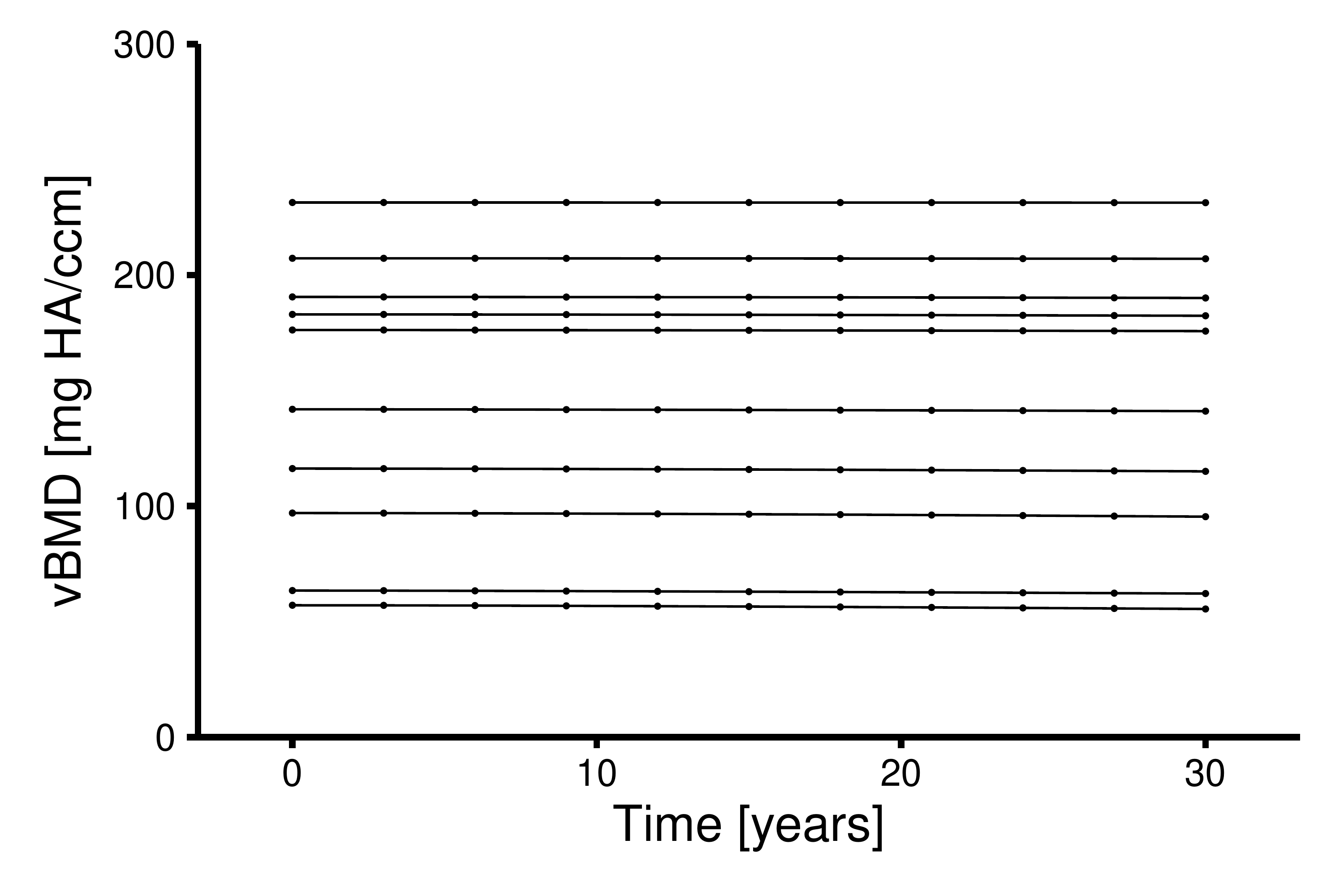}%
      \label{fig:morph:vbmd}
    } \\
    \subfloat[Conn.D]{
      \includegraphics[width=0.45\linewidth]{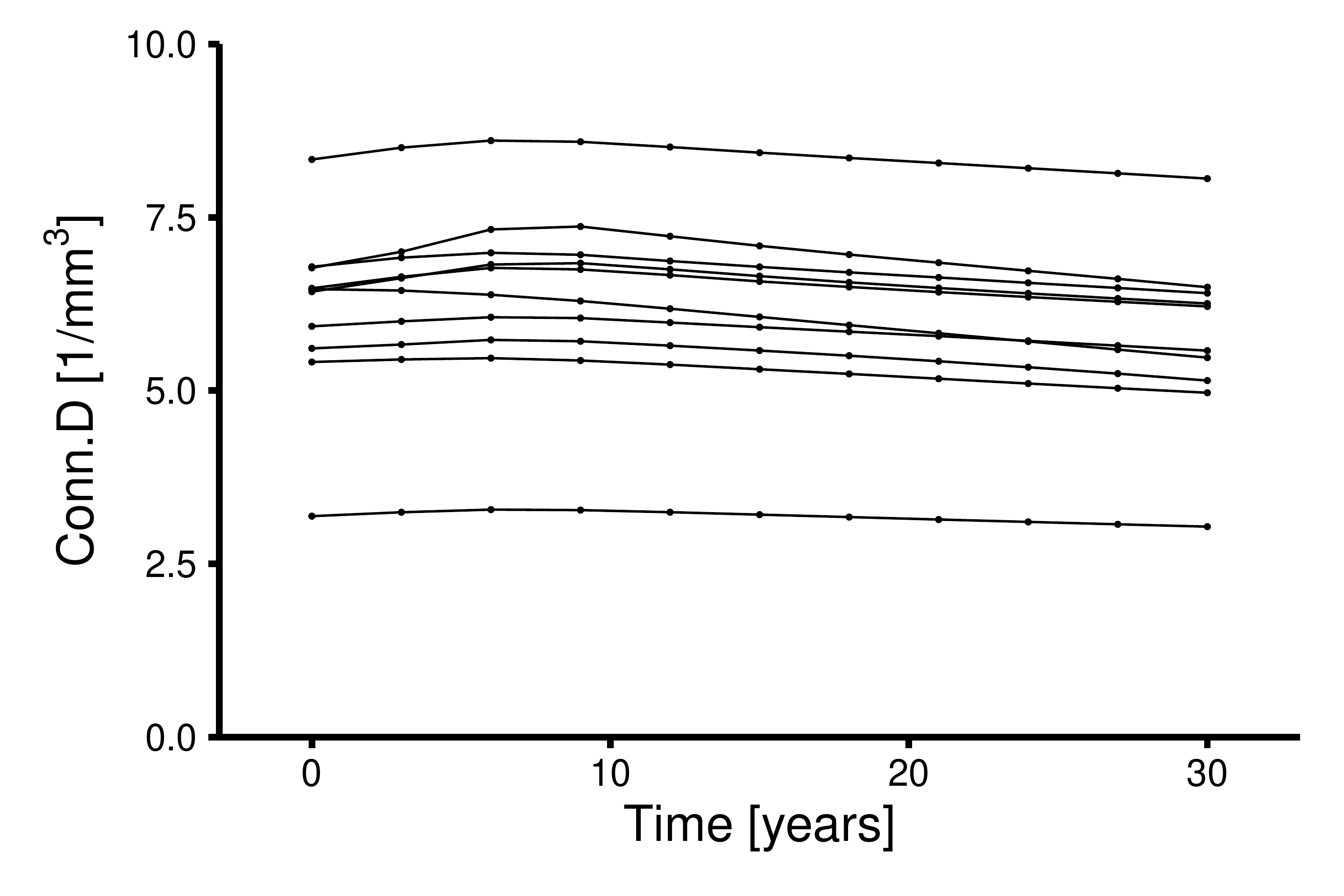}%
      \label{fig:morph:connd}
    } &
    \subfloat[SMI]{
      \includegraphics[width=0.45\linewidth]{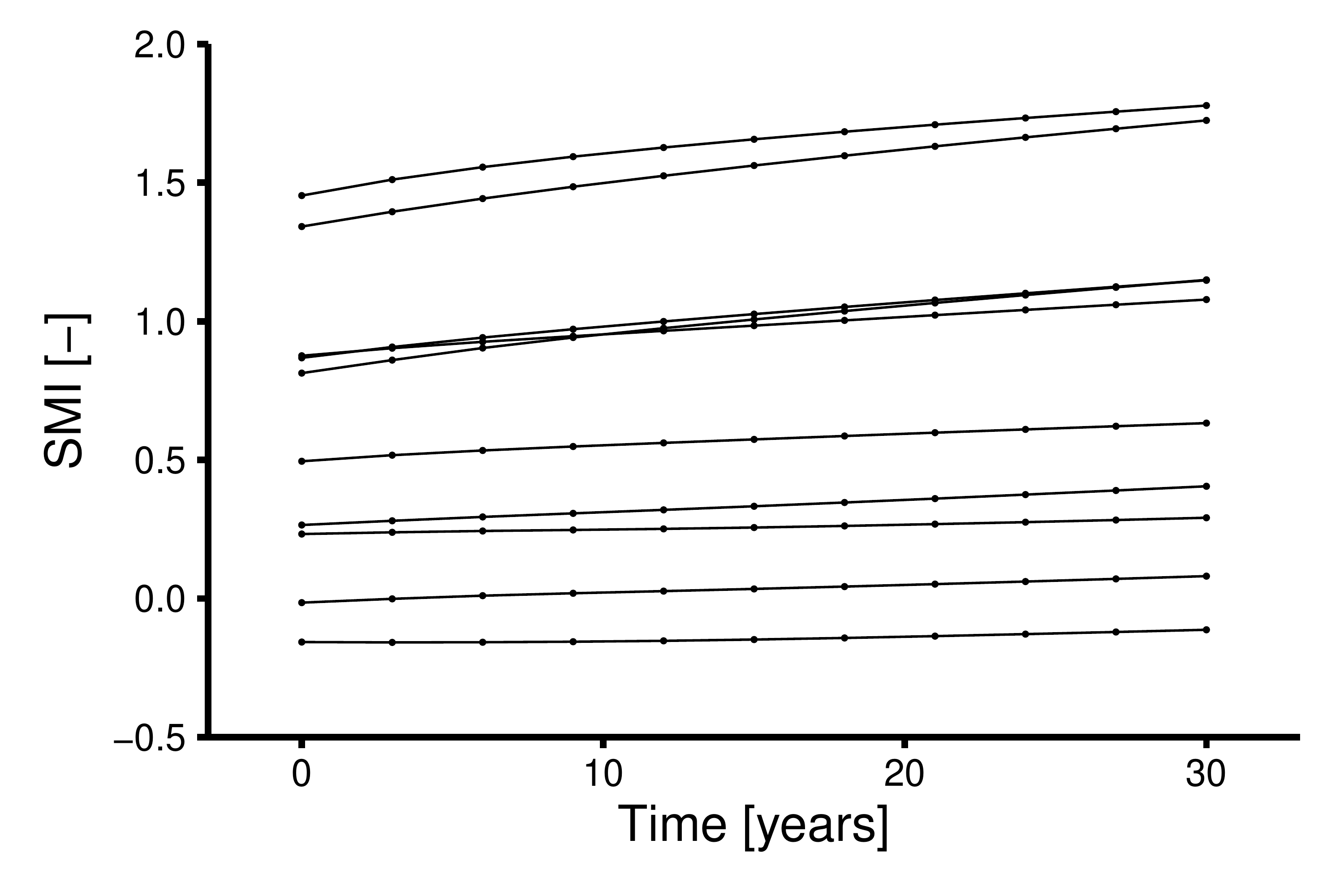}%
      \label{fig:morph:smi}
    }
  \end{tabular}
  \caption{Measured morphometry during the volume-preserving flow. Bone (\ref{fig:flow:bsbv}) surface area to volume ratio, (\ref{fig:flow:vbmd}) volumetric minearl density, (\ref{fig:flow:connd}) connectivity density, and (\ref{fig:flow:smi}) structure model index are plotted every three years over 30 years of simulation. Volumetric bone mineral density does not change as expected from the model. Connected lines are individual subjects.}
  \label{fig:morph}
\end{figure*}

\begin{figure*}
  \centering
  \begin{tabular}{ccc}
    \subfloat[Initial]{
      \includegraphics[width=0.3\linewidth]{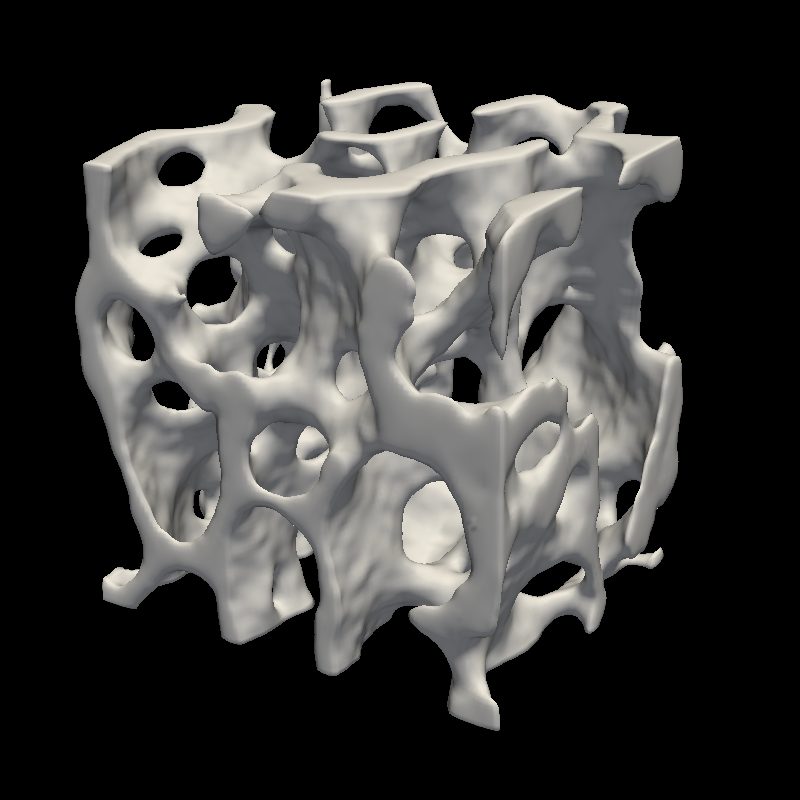}%
      \label{fig:image:0}
    } &
    \subfloat[15 Years]{
      \includegraphics[width=0.3\linewidth]{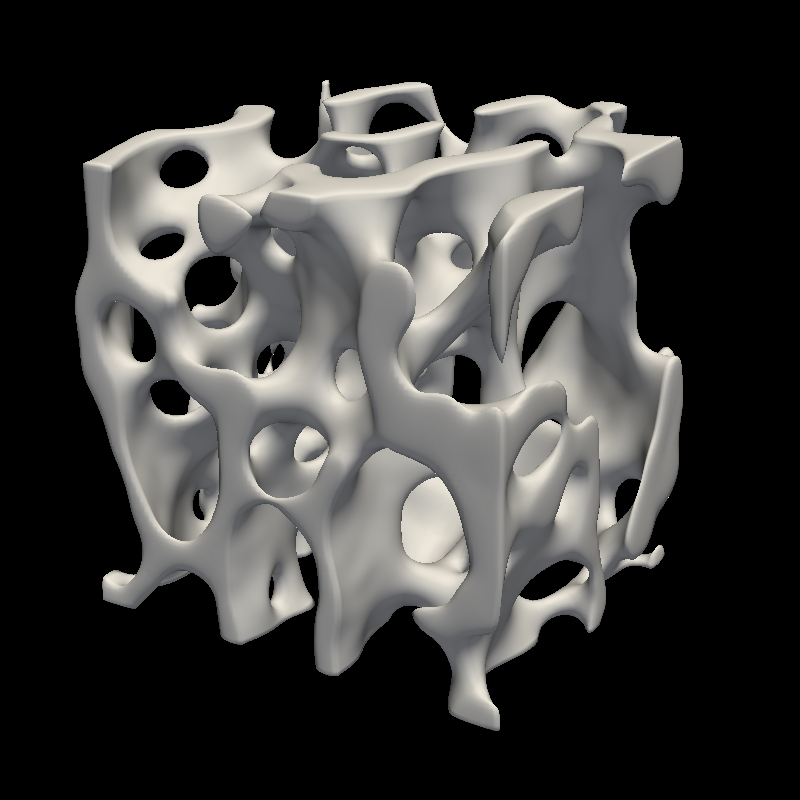}%
      \label{fig:image:15}
    } &
    \subfloat[30 Years]{
      \includegraphics[width=0.3\linewidth]{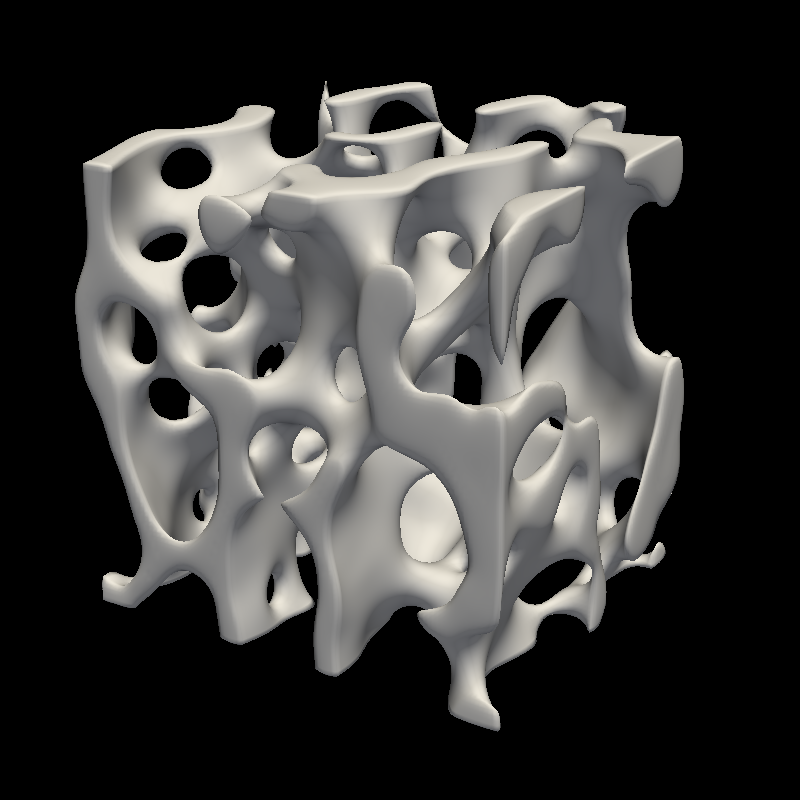}%
      \label{fig:image:30}
    }
  \end{tabular}
  \caption{Visualization of the surface during volume-preserving flow at three epochs. Rods can be seen disconnecting, suggesting a decrease in mechanical competence.}
  \label{fig:image}
\end{figure*}

Morphometrics for the volume-preserving flow are plotted in Figure~\ref{fig:morph}.
As expected, bone mineral density decreases only slightly, a result seen in a previous study~\cite{peng1999pde}.
The ratio of surface area to volume decreases over time while structure model index increases.
This is driven by the inverse relationship between SMI and BS/BV where surface average mean curvature is constant in the volume-preserving flow.
Since volume is constant, the surface area must be decreasing, consistent with area being the first variation of volume.
Finally, connectivity density increases for approximately 10 years then starts to decrease.
This is consistent with particles forming, increasing the \epc~characteristic, followed by rods disconnecting.
The initial, 15 year, and 30 year epochs of the median density subject are rendered in Figure~\ref{fig:image}.
Structural changes are subtle, but thin rods can be observed disconnecting and negative curvature areas thickening.
As in the non-volume-preserving case, noise on the surface is rapidly removed.

\section{Discussion}
\label{sec:discussion}
Bone adaptation is presented as a geometric flow.
Curvature-based bone adaptation is presented as the continuous version of the discrete simulated bone atrophy method.
The geometric flow can be simulated using the level set method, which naturally handles topological changes.
Two parameter sets were investigated, one resembling age-related bone loss and another being a volume-preserving flow.


The concretized model of curvature-based bone adaptation, and its predecessor simulated bone atrophy~\cite{muller1996analysis}, are unlikely to accurately predict \textit{in vivo} bone microarchitectural changes.
This owes to the relationship of the models to the Young-Laplace equation of surface tension, giving bubble-like architecutres if the model runs longer than 30 years.
However, the power of these models is in their clarity and computational abilities.
As with simulated bone atrophy, topological changes can be handled naturally.
Furthermore, the methods of simulated bone atrophy were central to developing a load-driven model that could handle topological changes~\cite{schulte2013strain}.
In this way, the specific instantiation (curvature-based bone adaptation) and general theory (bone adaptation as a geometric flow) should be separated~\cite{dirac1963evolution}.

Beyond a specific instantiation, there are limitations to describing bone adaptation as a geometric flow.
The central assumption to treat bone as a geometric flow is that the bone is orientable and smooth.
This assumption cannot be met during fetal development and fracturing healing where mineralization processes are not occurring on an existing surface.
Generally, ontogenesis and fracturing healing will be well described by internal remodeling methods~\cite{carter1984mechanical} while modeling and remodeling will be well described by external remodeling methods.
Such a situation alludes to a description that is not internal nor external remodeling.
There should be an underlying process which gives rise to internal or external remodeling based on circumstance.

In a search for the underlying dynamics of bone biology, a natural pattern similar to trabecular bone was sought.
An astonishing similarity is seen between trabecular bone and Turing patterns~\cite{turing1952chemical}.
Turing patterns emerge from simple reaction-diffusion equations, giving wonderfully complex shapes.
The Gray-Scott model is arguably the most studied of these models~\cite{gray1984autocatalytic} and extensive work has been done to classify the patterns as a function of their parameters~\cite{pearson1993complex}.
Another famous reaction-diffusion model is the Allen-Cahn equation~\cite{allen1979microscopic}, which was shown to converge to mean curvature flow~\cite{ilmanen1993convergence}.
Understanding trabecular patterning as a consequence of the dynamics of biochemistry will provide a deeper understanding of the multi-scale link in bone biology~\cite{webster2011silico}.

The obvious next step is to incorporate the level set method into a functional adaptation model.
Developing a functional adaptation model using the signed distance embedding is a concatenation of the biphasic model~\cite{besler2021constructing} and existing load-driven models~\cite{huiskes2000effects,adachi2001trabecular,schulte2013local}.
The key distinction is to perform finite element analysis on the density image constructed from the biphasic solution while having motion of the bone surface on the embedding.
This has the major advantage of not having to perform connected components filtering for finite element analysis, where isolated parts of bone relevant for measuring changes in total calcium must be removed in order to permit a solution of the finite element model.

Without binarizing the volume, it is difficult to perform connected components filtering during embedding or adaptation where isolated bone tissue components can arise.
Connected components filtering is needed to meet the assumptions of Odgaard in measuring connectivity density~\cite{odgaard1993quantification}.
The method presented here cannot assess the Betti numbers directly, but connectivity density is still estimated so the numerics are interpretable.
However, alternative methods exist for computing the Betti numbers of an implicit curve~\cite{pascucci2002efficient}, which would allow the direct measurement of $\beta_1$ without connected components filtering.
Furthermore, this method may be less sensitive than the total Gaussian curvature method used here.
Together with performing finite element analysis on the constructed density image, this would permit the development of disconnected components during adaptation, which would be important for monitoring calcium homeostasis.
Odgaard identified the issue of isolated bone particles during adaptation in his connectivity density work~\cite{odgaard1993quantification}:
\begin{quote}
  During bone formation and bone healing isolated islands of bone may exist, but these and related exceptions will not be considered further.
  The main reason for neglecting these exceptions is that fully isolated bony strands do only contribute very little, if at all, to the mechanical competence of a cancellous bone region.
\end{quote}

A major question that remains to be resolved is how to incorporate a local strain field and a local advection force together.
Existing models assume the magnitude of the diffeomorphism caused by the strain field is much smaller in magnitude than local changes in bone morphometry, consistent with the small-strain assumption of mechanics.
It is not obvious how to validate this assumption, nor how to incorporate the two types of motion together.
However, incorporating the strain field with the level set equation would provide a clear path to model dynamic behavior~\cite{turner1998three}, a currently under-modeled aspect of bone adaptation.

Lastly, curvature-based bone adaptation is intricately linked to surfaces of constant mean curvature.
These theories present a relationship between a functional being minimized (a Lagrangian) and a surface minimizing that Lagrangian, suggesting that bone can be a minimal surface of a measure different from curvature.
The concept of bone being optimal in some sense dates back to Wolff, with early formalization of the problem using Lagrangians and optimal control theory dating to Carter~\cite{jacobs1997adaptive}.
Early work on topology optimization given prescribed mechanical competence demonstrates structures remarkably similar to the cross-section of the humeral head~\cite{sethian2000structural}.
Deriving Frost's mechanostat --- the adaptation function, $F$ --- as the Euler-Lagrange of a Lagrangian may finally answer Huiskes~\cite{huiskes2000if}: If bone is the answer, then what is the question?





%

\appendix[Corrections to Odgaard]
\label{app:odgaard_corrections}
Error in Odgaard’s work on connectivity density~\cite{odgaard1993quantification} is straightforward to assess.
It arises from the difference in dimensionality of the surface, where Odgaard’s work did not modify the equations for \epc~characteristic between two-dimensional and three-dimensional structures.
The error can be seen immediately in Odgaard’s writing where he states~\cite{odgaard1993quantification}:

\begin{quote}
  For any solid body which can be deformed into a solid sphere, the Euler characteristic is 1; for any body which can be deformed into a solid torus, the Euler characteristic is 0, and generally
  \begin{equation*}
    \chi(v) = 1 - n
  \end{equation*}
  for a solid sphere with n handles.
\end{quote}

\noindent
This is true in two dimensions but the equation 
\begin{equation}
  \chi = 2 - 2g
\end{equation}
is appropriate in three dimensions, with $g$ the surface genus.
This oversight permeates the computation.
The correction then is to multiply the calculated \epc~characteristic by two, which corresponds roughly to doubling the connectivity density for large values of $\chi$.
Similarly, considering the object is solid, $\beta_2$ should be set to 1, but this is a minor correction given the highly negative \epc~characteristic of trabecular bone.

One can easily assess their morphometry software for this flaw by performing connectivity density on a three-dimensional sphere and torus, observing the \epc~characteristic.
The ideal results should be 2 for the sphere and 0 for the torus.
Software implemented for two dimensions will return 1 and 0.
A double-handled torus would return an Euler–Poincaré characteristic of -1 when it should return -2.

There is no obvious way to correct this mistake in the literature.
Many papers have measured bone as having half its true connectivity density.
A one-time backwards incompatible change to the field would cause confusion.
The authors recommend the field continues as-is with a note in the margins about bone actually being twice as connected as measured.





\bibliographystyle{IEEEtran}
\bibliography{IEEEabrv,Motion.bib}

\end{document}